\documentclass{article}

\usepackage{geometry}
\usepackage[utf8]{inputenc}
\usepackage[T1]{fontenc}
\usepackage[english]{babel}
\usepackage{fourier}

\usepackage{indentfirst}
\usepackage{setspace}
\usepackage{pgfornament}

\usepackage{amsmath, amsthm, amsfonts, amssymb}

\newtheorem{theorem}{Theorem}
\newtheorem{proposition}{Proposition}
\newtheorem{corollary}{Corollary}[theorem]
\newtheorem{lemma}{Lemma}[theorem]

\usepackage{booktabs}
\usepackage{multirow}

\definecolor{black}{HTML}{000000}
\definecolor{vermillion}{HTML}{D55E00}
\definecolor{orange}{HTML}{E69F00}
\definecolor{yellow}{HTML}{F0E442}
\definecolor{green}{HTML}{009E73}
\definecolor{sky_blue}{HTML}{56B4E9}
\definecolor{blue}{HTML}{0072B2}
\definecolor{purple}{HTML}{CC79A7}

\usepackage{natbib}
\setcitestyle{authoryear,open={(},close={)}}

\usepackage{hyperref}
\hypersetup{
	colorlinks,
	citecolor=teal,
	linkcolor=teal,
	urlcolor=teal
}

\AtBeginDocument{}
\AtBeginDocument{}
\AtBeginDocument{}
\AtBeginDocument{}

\makeatletter
\def\@footnotecolor{black}
\define@key{Hyp}{footnotecolor}{%
 \HyColor@HyperrefColor{#1}\@footnotecolor%
}
\patchcmd{\@footnotemark}{\hyper@linkstart{link}}{\hyper@linkstart{footnote}}{}{}
\makeatother

\title{How shared pedigrees shape gene genealogies in finite admixed populations}

\author{Gaspard Dousson Lys\(^1\), Camille Coron\(^1\), Tristan Mary-Huard\(^{1,2}\) and Paul Verdu\(^3\)}

\date{\footnotesize
	\(^1\)UMR MIA Paris-Saclay, AgroParisTech, INRAE, Université Paris-Saclay, 22 place de l'Agronomie, 91120 Palaiseau, France;
	\\
	\(^2\)UMR GQE - Le Moulon, AgroParisTech, CNRS, AgroParisTech, Université Paris-Saclay, 12 route 128, 91190 Gif-sur-Yvette, France;
	\\[-1mm]
	\(^3\)UMR7206 Eco-anthropology, CNRS, MNHN, Université Paris Cité, 57 rue Cuvier, 75005 Paris, France.
}

\begin{document}

\maketitle

\begin{abstract}
	Gene genealogies along a chromosome are dependent not only because nearby alleles tend to experience no recombination and are therefore inherited together, but also because alleles at different loci may share parental histories through a common ancestor. Disentangling physical linkage and shared parental trajectories is key to understanding how demography shapes genetic diversity. We develop a mathematical framework for describing the joint genealogies of sampled alleles in a finite admixed population, tracing their ancestry backward to the source populations while explicitly accounting for parental relationships and recombination events.

    Under a Wright–Fisher model, we show that the joint gene genealogy forms an absorbing Markov chain, derive the likelihood of local ancestry, and connect it to other quantities of interest, notably the admixture fraction. We further establish that, once demographic parameters become constant, local ancestry converges to distinct limiting distributions depending on the demographic setting. Under isolation, chromosomes undergo infinitely many recombination events and linkage disequilibrium is due to drift within population replicates, retaining traces of past admixture events. Under constant gene flow from source populations, recent contributions progressively erase the influence of earlier ancestry. Using the joint genealogy of two alleles, we obtain the limiting variance of admixture fractions in both regimes.

    We illustrate our framework’s flexibility through simulations of endogamy and selection. These examples show how genetic diversity depends on the structure of underlying parental relationships. Our work provides a framework for specifying and analyzing complex demographic scenarios and clarifies the links between local ancestry and mating patterns.
\end{abstract}

\section{Introduction}

In sexually reproducing species, individuals reproduce in pairs, yielding genealogies whose number of ancestors grows exponentially with time. By contrast, the number of crossover breakpoints in a given diploid genome, and thus the number of ancestors who contribute genetically to it, increases only linearly with the number of generations. Consequently, in large populations, not all genealogical ancestors can be genetic ancestors \citep{bairdDistributionSurvivingBlocks2003,gravelExistenceAbundanceGhost2015}. Nonetheless, the inference of kinship from genetic data has been successfully addressed, most notably through runs of homozygosity \citep{pembertonGenomicPatternsHomozygosity2012,ceballosRunsHomozygosityWindows2018}, as illustrated by the reconstruction of Punic kinships from ancient DNA by \cite{ringbauerPunicPeopleWere2025}. More generally, several works already established the significant influence of the \emph{pedigree}, here referring to what is sometimes called the population pedigree, namely the set of all parental relationships within a population, on genetic diversity \citep{ballGeneGenealogiesOrganismal1990,bartonRelationReproductiveValue2011,wakeleyGeneGenealogiesFixed2012,wakeleyEffectsPopulationPedigree2016,ralphEmpiricalApproachDemographic2019,diamantidisBurstsCoalescencePopulation2024}.

At large scales, pedigrees are structured rather than homogeneous, for instance due to geographical isolation. When focusing on roughly homogeneous subparts of a pedigree, certain demographic processes can disrupt genealogies by connecting previously isolated lineages. Such gene flow (\textit{e.g.}, migration) between previously isolated \emph{source populations} is called \emph{admixture}. Admixture models have been found widely applied, from crop species such as maize \citep{brandenburgIndependentIntroductionsAdmixtures2017} to human history \citep{hellenthalGeneticAtlasHuman2014}. Among anthropological cases, the trans-Atlantic slave trade is particularly well studied: between the 16th and 19th centuries, more than twelve millions Africans were enslaved and deported \citep{eltisAtlasTransatlanticSlave2010}, primarily to the Americas, but also to various regions of Africa and Europe. These forced migrations gave rise to a rich diversity of local admixture histories, as illustrated by the heterogeneous trajectories observed within the Cabo Verde archipelago \citep{laurentGeneticLinguisticAnalysis2023} or the nested admixture identified in São Tomé \citep{ciccarellaNestedAdmixtureTransAtlantic2025}. This historical period and its consequences provide a reference for our work, in which the onset of admixture is assumed known and only present-day genetic data are considered observable.

Admixed populations have been extensively studied in terms of population-level parameters such as contribution rates \citep{verduGeneralMechanisticModel2011,gravelPopulationGeneticsModels2012} and population sizes \citep{liangUnderstandingAdmixtureFractions2024}. However, these frameworks commonly rely (sometimes implicitly) on biparental Wright-Fisher mating and on averages over all possible pedigrees. Although \cite{tyukinQuenchedLimitsCoalescents2015} established that the convergence of coalescent processes towards the standard Kingman’s coalescent \citep{kingmanCoalescent1982,kingmanGenealogyLargePopulations1982} extends to settings conditioned on a fixed biparental Wright-Fisher pedigree, averaging over pedigrees can nevertheless fail to capture the dynamics of populations that deviate from standard model assumptions \citep[\textit{e.g.},][]{diamantidisBurstsCoalescencePopulation2024}, as human populations often do. To our knowledge, only a few recent studies directly address the individual-level mating patterns underlying population-level admixture histories. One approach models mating patterns as a genetically driven process, known as \emph{assortative mating}, in which the probability that two individuals mate depends on the similarity of their genomes with respect to ancestry from the source populations \citep{versluysWhyWePick2021,masseyDifferentiatingMechanismOutcome2025}. A second approach, derived from the model of \cite{verduGeneralMechanisticModel2011} for admixture histories, reformulates this problem in terms of combinatorics on ancestors \citep{mooneyNumberGenealogicalAncestors2023,agranat-tamirCountingGeneticAncestors2024,agranat-tamirSharedAncestorsBirthday2026}.

Even though the genealogical past cannot be fully reconstructed from present-day genetic data, genomic sequences are expected to retain signatures of the past evolutionary forces, such as drift and migration \citep[\textit{e.g.},][]{gutenkunstInferringJointDemographic2009}, selection \citep[\textit{e.g.},][]{fanGoingGlobalAdapting2016}, and culture \citep[\textit{e.g.},][]{heyerCulturalTransmissionFitness2005}. To simplify the nature of genetic data, we assume dealing with \emph{local ancestries} rather than \emph{identities by state}, that is, the origins of the observed alleles in terms of source populations rather than nucleotide bases or haplotypes. Such origins can be inferred from sequences using \emph{chromosome painting} methods, for which numerous algorithms are available \citep{priceSensitiveDetectionChromosomal2009,lawsonInferencePopulationStructure2012,baranFastAccurateInference2012,brisbinPCAdmixPrincipalComponentsBased2012,maplesRFMixDiscriminativeModeling2013,salter-townshendFineScaleInferenceAncestry2019}. These painted data are still expected to capture substantial information about admixture histories: \cite{gravelPopulationGeneticsModels2012} demonstrated that the decay of continuous ancestry tract lengths retains information on the duration and intensity of admixture processes. \cite{lambertChromosomePaintingHow2021} further illustrated that chromosome paintings can be highly detailed, although their study focused on identity by descent with respect to a single ancestor rather than local ancestry with respect to an entire source population.

In this article, we propose a flexible framework for modeling and studying demographic scenarios and mating patterns. We first introduce the model, which encodes all parental relationships in a graph, the pedigree, and describes the evolution of genetic diversity along it, simplified as chromosome paintings. We then consider the classical biparental Wright-Fisher mating and compute the probabilities of observing a given painting in the present generation by tracing back gene genealogies. In particular, we derive results for admixture fractions, that is, the proportion of an individual’s genome inherited from a given source population. Finally, we illustrate the flexibility of our approach through simulations of toy demographic models for endogamous populations and selective pressures, and discuss the generality of our results as well as directions for extending this framework to empirical data analyses.

\section{Model}

We characterize admixture histories as the interplay of three distinct yet intricately intertwined processes operating at different scales, each shaping the genetic diversity observed today. At the population level, admixture reflects gene flow from source populations into an admixed population. At the individual level, mating patterns determine the parental relationships through which genomes are transmitted across generations. At the genetic level, recombination gradually erodes, but never fully eliminates, physical linkage disequilibrium. We aim to precisely describe how past demographic processes acting between the population and individual scales have shaped the genetic diversity of present-day admixed individuals.

\subsection{Demography}

We consider a \emph{biparental Wright–Fisher model with admixture}, obtained by explicitly combining a Wright–Fisher demographic framework with the mechanistic admixture model of \cite{verduGeneralMechanisticModel2011}. The first generation (\(t = 1\)) of the admixed population is founded by individuals drawn from two reproductively isolated populations, denoted \(\textcolor{vermillion}{S_1}\) and \(\textcolor{blue}{S_2}\), in respective proportions \(\textcolor{vermillion}{s_1(0)}\) and \(\textcolor{blue}{s_2(0)}\), hereafter referred to as \emph{contributions}. In subsequent generations (\(t > 1\)), the source populations contribute proportions \(\textcolor{vermillion}{s_1(t-1)}\) and \(\textcolor{blue}{s_2(t-1)}\), with the remaining fraction, denoted \(h(t-1)\) such that \(h(t-1) + \textcolor{vermillion}{s_1(t-1)} - \textcolor{blue}{s_2(t-1)} = 1\), of parents drawn from the previous generation of the admixed population according to Wright–Fisher mating. The admixed population has a given finite reproductive size \(N_t\), possibly varying across generations \(t\), whereas the source populations are assumed infinite. Consequently, each drawn source individual is unique and unrelated to others, so that genetic drift occurs exclusively within the admixed population. Although admixture processes may involve more than two source populations, the extension to an arbitrary number of sources is straightforward and detailed in \autoref{app: arbitrary number of source populations}.

\begin{figure*}
	\centering
	\includegraphics{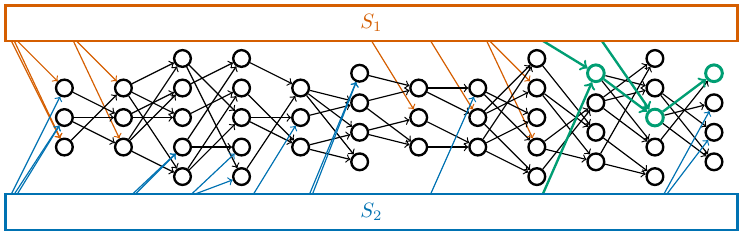}
	\caption{\textbf{Pedigree representation of parental relationships underlying an admixture history across generations.} \\
	Oriented edges are shown over \(\tau = 12\) generations for an admixed population whose sizes \((N_t)_{t \leq \tau}\) range from \(3\) to \(5\). Red arrows denote contributions from the source population \(\textcolor{vermillion}{S_1}\), blue arrows denote contributions from the source population \(\textcolor{blue}{S_2}\), and black arrows represent parent–offspring links within the admixed population itself. Together, these edges define the full genealogical graph, \textit{i.e.}, the pedigree. The genetic ancestors of a focal admixed individual’s maternal chromosome are highlighted in green.}
	\label{fig: pedigree}
\end{figure*}

To incorporate demographic structure such as mating patterns, we explicitly describe parental relationships between admixed individuals. Each parental relationship is an oriented edge to an admixed individual in generation \(t+1\) from either a source population or an admixed individual in generation \(t\). A \emph{pedigree} is a collection of parental relationships such that every admixed individual has exactly two distinct parents (\autoref{fig: pedigree}). Although the specific pedigree underlying an admixture history is typically unobservable, our goal is to characterize its impact on genetic diversity and to gain insight into how information about it can be inferred from genetic data.

In the biparental Wright-Fisher model with admixture, a random pedigree \(\Pi\) is defined as follows. For each admixed individual in generation \(t+1\), each parent is drawn independently from the source population \(\textcolor{vermillion}{S_1}\) with probability \(\textcolor{vermillion}{s_1(t)}\), from \(\textcolor{blue}{S_2}\) with probability \(\textcolor{blue}{s_2(t)}\), and from the admixed population otherwise, with probability \(h(t) = 1 - \textcolor{vermillion}{s_1(t)} - \textcolor{blue}{s_2(t)}\). When a parent is drawn from the admixed population, it is chosen uniformly at random; we recall that the two parents of a given individual are distinct, so that if both parents are drawn from the admixed population, the second parent is sampled uniformly from the remaining individuals.

While this null admixture model is entirely specified by population-level parameters, namely, source contributions and population sizes, the central feature of our framework, in contrast to others, is the pedigree, which encodes individual-level demography, such as mating patterns. This change in perspective allows the investigation of more complex demographic scenarios \citep[\textit{e.g.},][]{diamantidisBurstsCoalescencePopulation2024} by endowing \(\Pi\) with a parameterized probability law that incorporates additional meaningful parameters. Examples of such extensions, including endogamous mating structures and selective pressures, are discussed later.

\subsection{Recombination}

Genomes are inherited from parents, \textit{i.e.}, along the edges of the pedigree \(\Pi\), and thus retain information about its structure. An explicit model of recombination provides a framework for characterizing the genetic diversity arising from \(\Pi\), disentangling the effects of different demographic processes, and assessing what aspects of \(\Pi\) may be inferred from sequencing the present generation.

Each diploid individual carries a pair of chromosomes, \textit{i.e.}, two aligned sequences of alleles, represented over the segment \([0, \Lambda]\). Each locus \(\lambda\) carries two alleles, each labeled by its origin \(\textcolor{vermillion}{S_1}\) or \(\textcolor{blue}{S_2}\), indicating local ancestry. Formally, an individual is represented by an ancestry function
\begin{equation}
	\label{equ: ancestry function}
	A: [0, \Lambda] \to \big\{ \textcolor{vermillion}{S_1}, \textcolor{blue}{S_2} \big\}^2
\end{equation}
mapping loci to the origins of corresponding alleles on (arbitrarily ordered) maternal and paternal chromosomes. We recall that, in practice, local ancestry is inferred from sequenced identical-by-state alleles using chromosome painting methods.

\begin{figure*}
	\centering
	\includegraphics{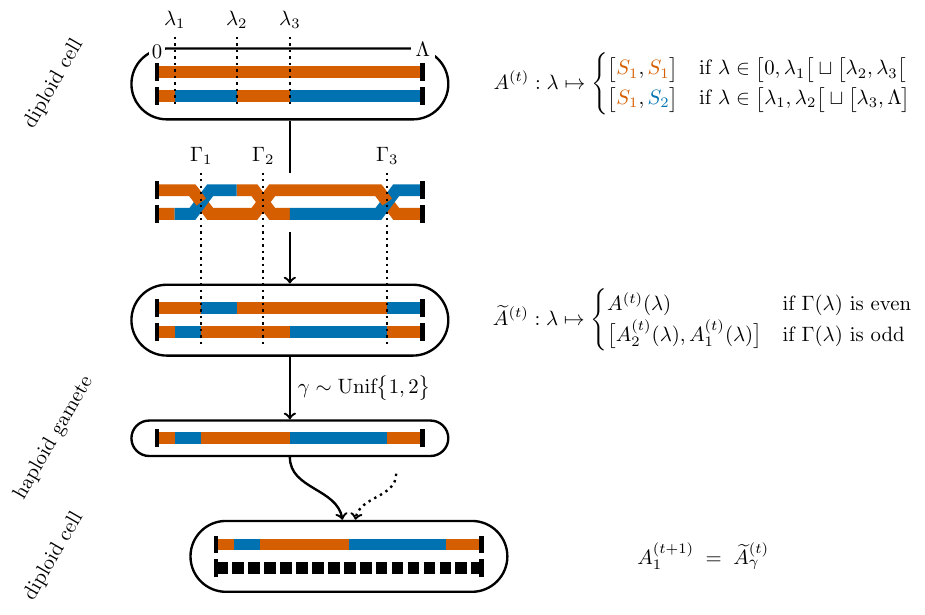}
	\caption{\textbf{Recombination is described as two successive and independent random processes.} \\
	Crossover events are modeled as ancestry switches occurring at loci drawn from a point process \(\Gamma\). Segregation is modeled as the random choice of one chromosome \(\gamma\) from each parental pair to form the offspring’s chromosome pair. Only the maternal lineage is shown here; the dotted paternal chromosome follows an identically distributed and independent process.}
	\label{fig: recombination}
\end{figure*}

Recombination is modeled by a point process \(\Gamma\) on the segment \([0, \Lambda]\) together with a random draw \(\gamma \in \{1, 2\}\) specifying which of the two chromomosomes is transmitted to the offspring. A realization of \(\Gamma\) determines crossover breakpoints along the chromosome pair; between successive breakpoints the ancestry function is switched. The coordinate \(\gamma\) is then inherited by the offspring (\autoref{fig: recombination}).

Under the model of \emph{perfectly flexible chromosomes}, \(\Gamma\) is a Poisson point process with intensity \(1\) on \([0, \Lambda]\) \citep{haldaneCombinationLinkageValues1919}, corresponding to measuring lengths in morgans. In practice, any base-pair sequence can be parametrized this way using empirical recombination maps \citep[\textit{e.g.},][]{kongHighresolutionRecombinationMap2002}. Nonetheless, we allow any point process \(\Gamma\) in what follows, provided that the number of breakpoints is almost surely finite, ensuring that the switches of ancestry function are well defined.

Consider alleles labelled \(1, 2, \ldots, n\) positioned at loci \(\lambda_1, \lambda_2, \ldots, \lambda_n\) in \([0, \Lambda]\). Assume these alleles reside on two homologous chromosomes, defining a partition \(\big(C'_1, C'_2\big)\) of \(\llbracket 1, n \rrbracket\), where \(C'_1\) and \(C'_2\) denote the sets of these alleles carried by the (arbitrarily ordered) first and second chromosomes. For any target partition \(\big(C_1, C_2\big)\), the probability
\[
	\rho_{C_1, C_2}^{C'_1, C'_2} \big(\lambda_1, \lambda_2, \ldots, \lambda_n \big)
\]
that recombination redistributes the alleles from \(\big(C'_1, C'_2\big)\) to \(\big(C_1, C_2\big)\) is fully determined by the crossover process \(\Gamma\). Of particular interest is the probability that a diploid parent with homologous chromosomes carrying the partition \(\big(C'_1, C'_2\big)\) transmits a recombined chromosome carrying precisely the alleles in \(C\). We denote this probability by
\[
	\rho_{C}^{C'_1, C'_2}
	\; = \;
	\rho_{C, \varnothing}^{C'_1, C'_2}.
\]

For example, if two alleles \(1\) and \(2\) lie on a perfectly flexible chromosome, then the probability that recombination separates them, \textit{i.e.}, that the number of crossover breakpoints between them is odd, is
\begin{equation}
	\label{equ: perfectly flexible recombination probability}
	\rho_{\{1\}, \{2\}}^{\{1, 2\}} \big( \lambda_1, \lambda_2 \big)
	\; = \;
	\frac{1 - \exp\big( - 2 \big| \lambda_2 - \lambda_1 \big| \big)}{2}.
\end{equation}
For short chromosome segments, local mechanical constraints invalidate the perfectly flexible approximation. Suppose instead that the chromosome is \emph{rigid}: crossing over occurs exactly once with probability \(\rho_0\) on \([0, \Lambda]\), and does not otherwise. Conditional on occurrence, the crossover breakpoint is drawn uniformly on the interval. In this case, the probability that recombination separates alleles \(1\) and \(2\) is
\begin{equation}
	\label{equ: rigid recombination probability}
	\rho_{\{1\}, \{2\}}^{\{1, 2\}} \big( \lambda_1, \lambda_2 \big)
	\; = \;
	\rho_0 \frac{\big| \lambda_2 - \lambda_1 \big|}{\Lambda}.
\end{equation}

\subsection{Forward process}

\begin{figure*}
	\centering
	\includegraphics{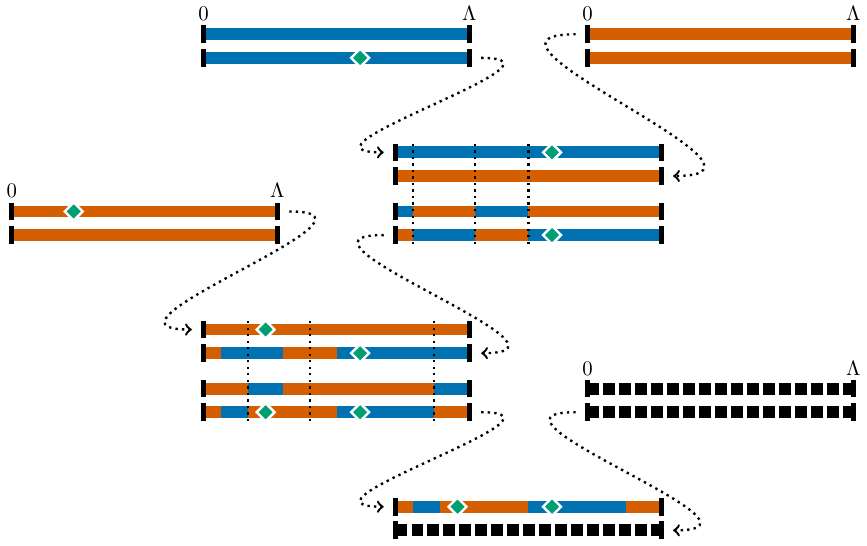}
	\caption{\textbf{Admixture history arises from the interplay of demographic and genetic events.} \\
	The genealogical tree of a focal individual’s maternal chromosome (bottom, highlighted in green in \autoref{fig: pedigree}) is traced back to source individuals as leaves (top). The paternal lineage is omitted for clarity (dotted chromosomes). Arrows indicate parental relationships across generations. Dotted lines represent recombination events, including both crossover locations \(\Gamma\) and chromosomal transmission choices \(\gamma\). The gene genealogies of the sampled alleles (green) must lie within, and are therefore constrained by, the genealogy of the entire chromosome, and ultimately by the pedigree. Recombination within source populations is not depicted, as it does not affect the local ancestry.}
	\label{fig: admixture history}
\end{figure*}

A realization of the pedigree together with all recombination events fully determines the admixed population at every generation. Although these processes operate on fundamentally different scales and are independent in nature, they interact intimately in an admixture history to generate complex genetic diversity. Starting from monochromatic chromosomes in the source populations, parental relationships are repeatedly drawn; chromosomes propagate through this pedigree structure and are progressively fragmented by crossovers (\autoref{fig: admixture history}). Over generations, this joint action of descent and recombination produces increasingly fine-grained mosaic genomes within the admixed population.

An admixture history is fundamentally a forward-in-time process. Yet, apart from very often highly degraded ancient DNA, we typically observe genetic variation only in the present generation, denoted \(\tau\). The complete ancestry function records, at every locus, the source population of each allele (\autoref{equ: ancestry function}). As already mentioned, this fine-scale information can be reasonably well inferred from sequencing data using chromosome painting methods. Assuming access to ancestry functions for a possibly small number of sampled admixed individuals, understanding how these observations reflect the underlying history, that is, how they are sensitive to the parameters of our model (\textit{e.g.}, source contributions, population sizes), requires adopting a backward-in-time perspective.

\subsection{Backward process}

When traced backward through recombination events and parental relationships, each allele eventually reaches the source population from which it originates (\autoref{fig: admixture history}). We refer to these backward paths as \emph{gene genealogies}, following \cite{coronPedigreeBiparentalMoran2022}. Because linkage disequilibrium couples the fates of nearby loci, these gene genealogies are dependent and must therefore be analyzed jointly.

Consider a set of \(n\) alleles sampled in generation \(\tau\) at loci \(\boldsymbol{\lambda} = \big( \lambda_1, \lambda_2, \ldots, \lambda_n \big)\), where loci may repeat because multiple alleles can be sampled at the same locus on different chromosomes. To jointly describe this sample in an ancestral generation, we must specify not only their genomic positions, or loci, but also their “demographic” positions: the ancestor and chromosome carrying them. Although the ancestral chromosomes themselves are unidentifiable, we can record which sampled alleles shared a chromosome at that time by introducing an equivalence relation under which two alleles are equivalent if they reside on the same chromosome. In addition, we define a second equivalence relation under which two chromosomes are equivalent if they are carried by the same individual. Together, these relations induce a two-level partition of \(\llbracket 1, n \rrbracket\) that encodes the alleles’ relative demographic positions. Moreover, in the context of chromosome painting, source populations are identified, so the corresponding classes are merged into a single labeled class for each source population. Otherwise, the representation is invariant under permutations of admixed individuals and transpositions of homologous admixed chromosomes.

Formally, let \(\mathcal{S}_2^{(n)}\) denote the collection of all (ordered) pairs of subsets of \(\llbracket 1, n \rrbracket\), one subset associated with the source population \(\textcolor{vermillion}{S_1}\) and the other with \(\textcolor{blue}{S_2}\). Let \(\mathcal{H}_N^{(n)}\) denote the collection of all sets of \(N\) unordered pairs of subsets of \(\llbracket 1, n \rrbracket\), one pair for each admixed individual, corresponding to its two homologous chromosomes. The product \(\mathcal{S}_2^{(n)} \times \mathcal{H}_N^{(n)}\) naturally maps to the collection of sets of \(2+2N\) subsets of \(\llbracket 1, n \rrbracket\) by forgetting both pairing and ordering. Such a set of \(2+2N\) subsets assigns each source population or admixed chromosome to alleles. However, each allele must be carried by exactly one chromosome, that is, must belong to exactly one subset. We therefore define \(\mathcal{P}_N^{(n)}\) as the inverse image under this mapping of all partitions of \(\llbracket 1, n \rrbracket\). Elements of \(\mathcal{P}_N^{(n)}\) correspond precisely to equivalence relations that can arise from a population with \(N\) admixed individuals. We also introduce
\[
	\mathcal{P}^{(n)}
	\; = \;
	\bigcup_{N \geq 0} \mathcal{P}_N^{(n)},
\]
the set of all possible allele configurations without limitations from the admixed population size.

\begin{figure*}
	\centering
	\includegraphics{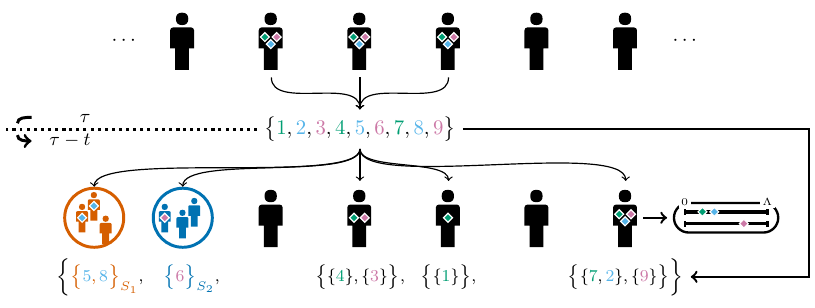}
	\caption{\textbf{Ancestor allele configurations are represented as partitions with identified source populations and paired chromosomes.} \\
	Nine alleles are sampled in the current generation \(\tau\) (top) in three admixed individuals on the three loci \(\textcolor{green}{0.58}\), \(\textcolor{sky_blue}{0.96}\) and \(\textcolor{purple}{1.93}\) of chromosome \(1\) (\(\Lambda \approx 2.7\)). Ancestors of these alleles in generation \(\tau - t\) are described by a partition (bottom), in which source populations are identified (left) among unlabeled admixed individuals. Chromosomes are grouped into unordered pairs (right).}
	\label{fig: allele configuration}
\end{figure*}

In generation \(t\), the ancestral configuration of the sample is represented by an element \(P \in \mathcal{P}_{N_t}^{(n)}\), which can be written as
\begin{equation}
	\label{equ: allele configuration}
	P
	\; = \; 
	\Big\{
		\textcolor{vermillion}{C_{S_1}}, \quad
		\textcolor{blue}{C_{S_2}}, \quad
		\big\{ C_{1, 1}, C_{1, 2} \big\}, \quad
		\big\{ C_{2, 1}, C_{2, 2} \big\}, \quad
		\ldots, \quad
		\big\{ C_{N_t, 1}, C_{N_t, 2} \big\}
	\Big\},
\end{equation}
where the labeling of admixed individuals and chromosomes within a pair is arbitrary, and where
\[
\llbracket 1, n \rrbracket
\; = \;
\textcolor{vermillion}{C_{S_1}} \sqcup \textcolor{blue}{C_{S_2}} \sqcup \bigsqcup_{i = 1}^N \big( C_{i, 1} \sqcup C_{i, 2} \big).
\]
Some subsets may be empty if the corresponding chromosome or source population carries no sampled allele; we sometimes omit such empty sets for brevity (as in \autoref{fig: allele configuration}). The configuration \(P\) has two components: \(\textcolor{vermillion}{C_{S_1}}\) and \(\textcolor{blue}{C_{S_2}}\) which encode the known source origins of the absorbed alleles at generation \(t\), and \(C_{1, 1}, C_{1, 2}, C_{2, 1}, \ldots\) which specify the admixed chromosomes carrying the other alleles at that generation. Note that \autoref{equ: allele configuration} naturally extends to any number of source populations, as detailed in \autoref{app: arbitrary number of source populations}.

Finally, the joint gene genealogy of the \(n\) sampled alleles at loci \(\boldsymbol{\lambda}\) is represented as a backward-in-time sequence \(G = \big(G^{(n)}_t(\boldsymbol{\lambda})\big)_{t \leq \tau}\) on \(\mathcal{P}^{(n)}\), terminating in \(\mathcal{S}_2^{(n)}\) by generation \(t = 0\) at the latest. Note that \(G^{(n)}_t(\boldsymbol{\lambda}) \in \mathcal{P}_{N_t}^{(n)}\) for every generation \(t\). In what follows, we suppress explicit dependence on the number of sampled alleles \(n\) and on the loci vector \(\boldsymbol{\lambda}\) whenever this does not cause ambiguity. While our representation differs from that of \cite{esserPartitioningDualityLinkage2016}, notably through the introduction of source populations, their framework served as an initial inspiration for \autoref{equ: allele configuration}.

To illustrate, consider the common situation in which \(n\) individuals in present generation \(\tau\) are sequenced at \(k\) loci along a given chromosome. This yields \(nk\) gene genealogies to be traced backward in time, or \(2nk\) if both homologous chromosomes are sequenced. For concreteness, suppose we sample \(n = 3\) distinct individuals and sequence one of their chromosomes 1, whose total length is approximately \(\Lambda \approx 2.7\) morgans \citep{kongHighresolutionRecombinationMap2002}, at \(k = 3\) arbitrary loci, \textit{e.g.}, \(0.58\), \(0.96\), and \(1.93\). We label the sampled alleles from \(1\) to \(9\): alleles \(1\)--\(3\) belong to the first individual, \(4\)--\(6\) to the second, and \(7\)--\(9\) to the third.

At the sampled generation \(\tau\), each allele is fully characterized by its locus together with its equivalence class under the following hierarchical partition:
\[
	\big\{ \{ 1, 2, 3 \} \big\}, \quad \big\{ \{ 4, 5, 6 \} \big\}, \quad \big\{ \{ 7, 8, 9 \} \big\},
\]
where inner braces denote alleles on the same sequenced chromosome 1, and outer braces denote chromosomes carried by the same sampled individual.

We now trace these alleles back \(t\) generations into the ancestry of the three sampled genomes. The loci remain fixed, but recombination and coalescence events may split or merge the equivalence classes defined above. A possible configuration in generation \(\tau - t\) is
\[
	\big\{ \{ 5 \} \big\}, \quad \big\{ \{ 8 \} \big\}, \quad \big\{ \{ 6 \} \big\}, \quad \big\{ \{ 4 \}, \{ 3 \} \big\}, \quad \big\{ \{ 1 \} \big\}, \quad \big\{ \{ 7, 2 \}, \{ 9 \} \big\}.
\]
In this partition, the nine alleles are distributed across eight ancestral chromosomes but only six ancestral individuals. Moreover, some of these individuals may belong to a source population rather than to the admixed population. For instance,
\[
	\big\{ \{ 5 \} \big\}, \quad \big\{ \{ 8 \} \big\}
	\; \to \;
	\textcolor{vermillion}{\big\{ 5, 8 \big\}_{S_1}}
	\quad \textrm{and} \quad
	\big\{ \{ 6 \} \big\}
	\; \to \;
	\textcolor{blue}{\big\{ 6 \big\}_{S_2}}.
\]
With this convention, a convenient representation of the demographic positions of the sampled alleles in generation \(\tau - t\) (\autoref{fig: allele configuration}) is
\[
	\textcolor{vermillion}{\big\{ 5, 8 \big\}_{S_1}}, \quad \textcolor{blue}{\big\{ 6 \big\}_{S_2}}, \quad \big\{ \{ 4 \}, \{ 3 \} \big\}, \quad \big\{ \{ 1 \} \big\}, \quad \big\{ \{ 7, 2 \}, \{ 9 \} \big\}.
\]

Such notations allow us to precisely track joint gene genealogies. For example, consider the genealogical tree of the chromosome shown in \autoref{fig: admixture history} and trace backward the two alleles \(1\) and \(2\) sampled in generation \(\tau\). The resulting gene genealogy can be expressed as the configuration sequence
\[
	\big\{ \{ 1, 2 \} \big\}
	\quad \to \quad
	\big\{ \{ 1 \}, \{ 2 \} \big\}
	\quad \to \quad
	\textcolor{vermillion}{\big\{ 1 \big\}_{S_1}}, \big\{ \{ 2 \} \big\}
	\quad \to \quad
	\textcolor{vermillion}{\big\{ 1 \big\}_{S_1}}, \textcolor{blue}{\big\{ 2 \big\}_{S_2}}.
\]
Initially, both alleles lie on the same chromosome and therefore share the same parent. Moving backward in time, a recombination event separates them. The resulting ancestral chromosomes must originate from different parents. One parent belongs to source population \(\textcolor{vermillion}{S_1}\), while the other allele remains in the admixed population for one additional generation. Finally, the joint genealogy stops once the second allele reaches source population \(\textcolor{blue}{S_2}\).

\section{Results}

As noted, only a small part of an admixture history is observable in practice. Accordingly, we assume sampling alleles from a few living individuals and inferring their origin correctly using chromosome painting methods. Sampling is formalized as the choice of loci and chromosomes to be sequenced, that is, by selecting a loci vector \(\boldsymbol{\lambda} \in [0, \Lambda]^n\) together with an allele configuration \(P \in \mathcal{H}_{N_\tau}^{(n)}\) for the present generation \(\tau\). Inferring local ancestries among these alleles amounts to observing the corresponding absorbing configuration \(P_0 \in \mathcal{S}_2^{(n)}\) ultimately reached by the joint gene genealogy \(G\) initiated from \(P\). All information obtained by chromosome sequencing and painting is thus captured by the sampled loci \(\boldsymbol{\lambda}\), the present allele configuration \(G_\tau = P\), and the corresponding ancestral origins given by \(G_0 = P_0\).

Although the full trajectory of the joint gene genealogy \(G\) between \(P\) and \(P_0\) is intractable, knowledge of the initial and terminal configurations already constrains the possible shapes of the parental relationships underlying such trajectories. We begin by studying the local transitions of \(G\). Conditional on a sampling configuration \(P\) in generation \(\tau\), we derive the probability distribution of the induced local ancestries, that is, the absorbing configuration \(P_0\) in generation \(t = 0\). This framework enables us to revisit the extensively studied admixture fraction and to assess how much information it can convey about the underlying pedigree \(\Pi\).

\subsection{Absorbing and Markov properties}

By the randomness of the pedigree \(\Pi\) and of the recombination events, the joint gene genealogy \(G\) is a stochastic process on the configuration space \(\mathcal{P}\). Let \(\Pi_t\) denote the restriction of \(\Pi\) to parental relationships between generation \(t\) and \(t+1\). Conditionally on the allele configuration \(G_{t+1}\), the pedigree layer \(\Pi_t\), and the recombination events at generation \(t\), the configuration one generation earlier \(G_t\) is fully determined: \(G\) satisfies a backward recursion. Recombination events are independent across generations, and in the biparental Wright-Fisher model with admixture the pedigree layers are independent as well, so that \(G\) is a Markov chain \citep{kallenbergFoundationsModernProbability2021}.

Since the distribution of \(\Pi_t\) depends explicitly on the generation \(t\) through the population sizes \(N_t\) and \(N_{t+1}\) and through the contributions \(\textcolor{vermillion}{s_1(t)}\) and \(\textcolor{blue}{s_2(t)}\), the Markov chain \(G\) is inhomogeneous. Marginal genealogies are progressively absorbed into the source populations until the joint genealogy reaches a configuration in \(\mathcal{S}_2\), which occurs during the foundation generation (for \(t = 0\)) to the latest case; therefore, the Markov chain \(G\) is absorbing.

To describe transitions of \(G\), we first identify parental allele configurations \(P' \in \mathcal{P}_{N_t}\) that are reachable from a given configuration \(P \in \mathcal{P}_{N_{t+1}}\). Using the notation introduced in \autoref{equ: allele configuration} for \(P\) (and primes for \(P'\)), three conditions ensure accessibility:
\begin{enumerate}
	\item \textbf{Absorption in source population:}
	alleles already absorbed in \(P\) were already absorbed in \(P'\):
		\[
			\textcolor{vermillion}{C_{S_1}}
			\; \subset \;
			\textcolor{vermillion}{C'_{S_1}},
			\quad
			\textcolor{blue}{C_{S_2}}
			\; \subset \;
			\textcolor{blue}{C'_{S_2}}.
		\]
		Note that this follows from the assumed reproductive isolation of source populations, whose relevance and limitations are examined in the Discussion.
	\item \textbf{Chromosome inheritance:}
	each chromosome of an admixed individual in \(P\) must have been inherited from a single parent, either a source population or an admixed individual:
		\[
			\forall i \leq N_{t+1}, \forall \delta \in \{ 1, 2 \},
			\quad
			C_{i, \delta}
			\; \subset \;
			\textcolor{vermillion}{C'_{S_1}}
			\quad \textrm{or} \quad
			C_{i, \delta}
			\; \subset \;
			\textcolor{blue}{C'_{S_2}}
			\quad \textrm{or} \quad
			\exists j \leq N_t, \;
			C_{i, \delta}
			\; \subset \;
			\Big( C'_{j, 1} \sqcup C'_{j, 2} \Big).
		\]
		Let \(\nu_{t+1}^{(\textcolor{vermillion}{S_1})}\), \(\nu_{t+1}^{(\textcolor{blue}{S_2})}\), and \(\nu_{t+1}^{(H)}\) denote the numbers of admixed chromosomes carrying sampled alleles in \(P\) directly inherited from source populations \(\textcolor{vermillion}{S_1}\) and \(\textcolor{blue}{S_2}\), and from the admixed population, respectively. Recombination in admixed parents may split and rejoin chromosomal segments, potentially separating allele equivalence classes backward in time.
	\item \textbf{Absence of selfing:}
	if both chromosomes of an individual in \(P\) come from the admixed population, they cannot originate from the same individual in \(P'\):
		\[
			\forall i \leq N_{t+1}, \forall j \leq N_t,
			\quad
			\Big( C_{i, 1} \cup C_{i, 2} \Big)
			\; \subset \;
			\Big( C'_{j, 1} \cup C'_{j, 2} \Big)
			\quad \implies \quad
			C_{i, 1}
			\; = \;
			\emptyset
			\quad \textrm{or} \quad
			C_{i, 2}
			\; = \;
			\emptyset.
		\]
		This introduces a small correction in transition probabilities due to dependence in parental draws. Let \(N_{t+1}^{(2,2)}\) denote the number of admixed individuals in \(P\) whose two homologous chromosomes carry sampled alleles and are inherited from two admixed parents.
\end{enumerate}
If all three conditions are satisfied, \(P'\) is accessible from \(P\); otherwise the transition probability must be zero under our model assumptions. Note that some admixed individuals in generation \(t\) may contribute no chromosome carrying sampled alleles, and let \(\widetilde{N}_t \leq N_t\) denote the number of admixed parents in \(P'\), that is, the number of individuals \(j\) for which at least one chromosome \(\delta\) satisfies \(C_{j, \delta} \not= \emptyset\). With all these notations, we can explicitly characterize the dynamics of gene genealogies with transition kernels (\autoref{thm: absorbing markov chain}, proof in \autoref{app: absorbing markov chain}).

\begin{theorem}
	\label{thm: absorbing markov chain}
	Let \(G\) denote the joint gene genealogy of \(n\) alleles at loci \(\boldsymbol{\lambda}\) under the biparental Wright-Fisher model with admixture. Then \(G\) is a time-reversed absorbing Markov chain, and its transition matrix from generation \(t+1\) to \(t\) is given by
	\begin{equation}
		\label{equ: transition matrix}
		Q_t(\boldsymbol{\lambda}) \big[ P \to P' \big]
		\; = \;
		\textcolor{vermillion}{s_1(t)}^{\nu_{t+1}^{(\textcolor{vermillion}{S_1})}} \textcolor{blue}{s_2(t)}^{\nu_{t+1}^{(\textcolor{blue}{S_2})}} \Bigg( \frac{h(t)}{2 N_t} \Bigg)^{\nu_{t+1}^{(H)}} \Bigg( \frac{N_t}{N_t-1} \Bigg)^{N^{(2, 2)}_{t+1}} \prod_{j = 1}^{\widetilde{N}_t} \Bigg( 2 \big(N_t - j + 1\big) \prod_{(i, \delta) \leftarrow j} \rho_{C_{i, \delta}}^{C'_{j, 1}, C'_{j, 2}}(\boldsymbol{\lambda}) \Bigg),
	\end{equation}
	where the product over \((i, \delta) \leftarrow j\) ranges over all chromosomes \((i, \delta) \in \big\llbracket 1, N_{t+1} \big\rrbracket \times \{1, 2\}\) carrying sampled alleles (\(C_{i, \delta} \not= \emptyset\)) and being inherited from the admixed parent \(j \in \big\llbracket 1, \widetilde{N}_t \big\rrbracket\), and where the transition is taken over any configuration \(P'\) accessible from \(P\). The recombination probability function $\rho$ is determined by the point process $\Gamma$, \textit{e.g.}, \autoref{equ: perfectly flexible recombination probability} and \autoref{equ: rigid recombination probability}.
\end{theorem}

By standard properties of absorbing Markov chains \citep{grinsteadIntroductionProbability1998}, these transition kernels \(\big(Q_t(\boldsymbol{\lambda})\big)_{t \leq \tau}\) can be used to derive the distribution of local ancestries at the sampled loci \(\boldsymbol{\lambda}\) directly from the model parameters. For each generation \(t\), we use the canonical block decomposition
\begin{equation}
	\label{equ: transition matrix decomposition}
	Q_t(\boldsymbol{\lambda})
	\; = \;
	\begin{bmatrix}
		\widetilde{Q}_t(\boldsymbol{\lambda}) & R_t(\boldsymbol{\lambda}) \\ \mathbf{0} & I_{2^n}
	\end{bmatrix},
\end{equation}
where \(\widetilde{Q}_t(\boldsymbol{\lambda})\) describes transitions among transient configurations, \(R_t(\boldsymbol{\lambda})\) describes transitions from transient to absorbing configurations, and \(I_{2^n}\) is the \(2^n \times 2^n\) identity matrix (describing transitions among absorbing configurations). We next introduce the \emph{absorption matrix}
\begin{equation}
	\label{equ: absorption matrix}
	M_\tau(\boldsymbol{\lambda})
	\; = \;
	\sum_{t = 0}^{\tau - 1} \widetilde{Q}_{\tau-1}(\boldsymbol{\lambda}) \widetilde{Q}_{\tau-2}(\boldsymbol{\lambda}) \cdots \widetilde{Q}_{t+1}(\boldsymbol{\lambda}) R_t(\boldsymbol{\lambda}),
\end{equation}
where the empty product is interpreted as the identity matrix. The row of \(M_\tau(\boldsymbol{\lambda})\) indexed by a transient configuration \(P\) gives the probability distribution over the absorbing configurations reached when the gene genealogy is initiated from \(P\) in generation \(\tau\). In other words, it describes the likelihood of chromosome painting outcomes conditional on sampling loci \(\boldsymbol{\lambda}\) on the chromosomes specified by \(P\) for the demographic parameters \(N_t\), \(\textcolor{vermillion}{s_1(t)}\), and \(\textcolor{blue}{s_2(t)}\) (implicitly fixed in the computation). More generally, for any integer \(p \geq 0\), we define the time-weighted absorption matrix of order \(p\) as
\[
	M^{(p)}_\tau(\boldsymbol{\lambda})
	\; = \;
	\sum_{t = 0}^{\tau - 1} (\tau-t)^p \widetilde{Q}_{\tau-1}(\boldsymbol{\lambda}) \widetilde{Q}_{\tau-2}(\boldsymbol{\lambda}) \cdots \widetilde{Q}_{t+1}(\boldsymbol{\lambda}) R_t(\boldsymbol{\lambda}).
\]

By explicitly characterizing the probability distribution over the absorbing configurations, the absorption matrix \(M_\tau(\boldsymbol{\lambda})\) captures how model parameters shape the ancestry of any finite sample of alleles. This information is sufficient to study moments of a wide range of summary statistics commonly used in admixture analyses. Among these, the extensively used \emph{admixture fraction} \citep{verduGeneralMechanisticModel2011,fortes-limaComplexGeneticAdmixture2021,liangUnderstandingAdmixtureFractions2024} quantifies the proportion of an individual’s genome inherited from a given source population.

For an admixed individual in generation \(\tau\), with ancestry function \(A^{(\tau)}\), the admixture fraction with respect to source population \(\textcolor{vermillion}{S_1}\) is
\begin{equation}
\label{equ: admixture fraction}
H^{(\tau)}_{\textcolor{vermillion}{S_1}}
\; = \;
\frac{1}{2 \Lambda} \int_{[0, \Lambda]} \mathbf{1}_{[A^{(\tau)}_1(\lambda) = \textcolor{vermillion}{S_1}]} + \mathbf{1}_{[A^{(\tau)}_2(\lambda) = \textcolor{vermillion}{S_1}]} \mathrm{d}\lambda.
\end{equation}
This statistic is the proportion of an individual’s genes inherited from the source population \(\textcolor{vermillion}{S_1}\). It therefore provides a coarse yet informative summary of the share of an individual’s ancestors coming from that population, collapsing the fine-scale ancestry mosaic into a single scalar. However, recall that an individual’s genome is not representative of its full genealogy: not all genealogical ancestors are genetic ancestors, and genetic ancestors do not contribute equally. Importantly, admixture fractions are invariant to the arbitrary ordering of chromosomes in the ancestry function.

\begin{proposition}
	\label{prop: admixture fraction moments}
	Let \(H^{(\tau)}_{\textcolor{vermillion}{S_1}}\) denote the admixture fraction with respect to source population \(\textcolor{vermillion}{S_1}\) of a uniformly sampled admixed individual in generation \(\tau\). Then, for any integer \(p \geq 0\), the \(p\)-th moment of \(H^{(\tau)}_{\textcolor{vermillion}{S_1}}\) is given by
	\begin{equation}
		\label{equ: admixture fraction moments}
		\mathbf{E} \Big[ \big( H^{(\tau)}_{\textcolor{vermillion}{S_1}} \big)^p \Big]
		\; = \;
		\frac{1}{\Lambda^p} \int_{[0, \Lambda]^p} \frac{1}{2^p} \sum_{k = 0}^p \binom{p}{k} M_\tau(\boldsymbol{\lambda}) \Big[ \Big\{ \big\{ \{ 1, \ldots, k \}, \{ k+1, \ldots, p \} \big\} \Big\} \to \Big\{ \textcolor{vermillion}{\big\{ 1, \ldots, p \big\}_{S_1}} \Big\} \Big] \mathrm{d}\boldsymbol{\lambda},
	\end{equation}
	where \(M\big[P \to P_0\big]\) denote the \(\big(P, P_0\big)\)-coefficient of the matrix \(M\).
\end{proposition}

\begin{proof}
	We recall that the admixture fraction \(H^{(\tau)}_{\textcolor{vermillion}{S_1}}\) is the proportion of alleles inherited from the source population \(\textcolor{vermillion}{S_1}\) in the sampled individual of generation \(\tau\) (\autoref{equ: admixture fraction}). Consider its \(p\)-th power. Expanding the product inside a unified integral,
	\[
		\begin{aligned}
			\big( H^{(\tau)}_{\textcolor{vermillion}{S_1}} \big)^p
			& \; = \;
			\frac{1}{(2 \Lambda)^p} \prod_{k = 1}^p \int_{[0, \Lambda]} \mathbf{1}_{[A^{(\tau)}_1(\lambda_k) = \textcolor{vermillion}{S_1}]} + \mathbf{1}_{[A^{(\tau)}_2(\lambda_k) = \textcolor{vermillion}{S_1}]} \mathrm{d}\lambda_k
			\\
			& \; = \;
			\frac{1}{(2 \Lambda)^p} \int_{[0, \Lambda]^p} \prod_{k = 1}^p \Big( \mathbf{1}_{\big[ A_1(\lambda_k) = \textcolor{vermillion}{S_1} \big]} + \mathbf{1}_{\big[ A_2(\lambda_k) = \textcolor{vermillion}{S_1} \big]} \Big) \mathrm{d}\boldsymbol{\lambda}.
		\end{aligned}
	\]
	Here, \(A\) denotes the ancestry function of the sampled individual whose admixture fraction is being computed, and \(\boldsymbol{\lambda} = (\lambda_1,\ldots,\lambda_p)\) represents \(p\) genomic positions that vary independently over \([0,\Lambda]\). Expanding the product into a sum over all choices of which chromosome carries each allele,
	\[
		\big( H^{(\tau)}_{\textcolor{vermillion}{S_1}} \big)^p
		\; = \;
		\frac{1}{(2 \Lambda)^p} \sum_{\delta_1 = 1}^2 \cdots \sum_{\delta_p = 1}^2 \int_{[0, \Lambda]^p} \prod_{k = 1}^p \mathbf{1}_{\big[ A_{\delta_k}(\lambda_k) = \textcolor{vermillion}{S_1} \big]} \mathrm{d}\boldsymbol{\lambda}.
	\]
	Since the integral is invariant under permutations of the coordinates of \(\boldsymbol{\lambda}\), the terms may be grouped according to the number \(k\) of loci assigned to the “first” chromosome:
	\[
		\big( H^{(\tau)}_{\textcolor{vermillion}{S_1}} \big)^p
		\; = \;
		\frac{1}{(2 \Lambda)^p} 
		\sum_{k = 1}^p \binom{p}{k} \int_{[0, \Lambda]^p}
		\mathbf{1}_{\big[ A_{1}(\lambda_1) = \textcolor{vermillion}{S_1} \big]} \cdots \mathbf{1}_{\big[ A_{1}(\lambda_k) = \textcolor{vermillion}{S_1} \big]} \mathbf{1}_{\big[ A_{2}(\lambda_{k+1}) = \textcolor{vermillion}{S_1} \big]} \cdots \mathbf{1}_{\big[ A_{2}(\lambda_p) = \textcolor{vermillion}{S_1} \big]} \mathrm{d}\boldsymbol{\lambda}.
	\]
	Taking expectations yields
	\[
		\mathbf{E} \big[ \big( H^{(\tau)}_{\textcolor{vermillion}{S_1}} \big)^p \big]
		\; = \;
		\frac{1}{(2 \Lambda)^p} \int_{[0, \Lambda]^p} \sum_{k = 1}^p \binom{p}{k} \mathbf{P} \Big[ \substack{ A_{1}(\lambda_1) = \cdots = A_{1}(\lambda_k) = \textcolor{vermillion}{S_1} \\ A_{2}(\lambda_{k+1}) = \cdots = A_{2}(\lambda_p) = \textcolor{vermillion}{S_1} } \Big] \mathrm{d}\boldsymbol{\lambda}.
	\]

	For fixed loci \(\lambda_1, \ldots, \lambda_p\), the event whose probability appears in the integrand corresponds to the absorption of the \(p\) marginal gene genealogies into the source population \(\textcolor{vermillion}{S_1}\), \textit{i.e.}, into the absorbing configuration
	\[
		P_0
		\; = \;
		\textcolor{vermillion}{ \big\{ 1, \ldots, p \big\}_{S_1}}
	\]
	given that at generation \(\tau\) the configuration of sampled alleles is
	\[
		P
		\; = \;
		\big\{ \{ 1, \ldots, k \}, \{ k+1, \ldots, p \} \big\}.
	\]
	By definition of the absorption matrix \(M_\tau(\boldsymbol{\lambda})\) for the joint gene genealogy \(G\) of alleles \(1, \ldots, p\) at loci \(\boldsymbol{\lambda} = \big( \lambda_1, \ldots, \lambda_p \big)\),
	\[
		\mathbf{P} \big[ G_0 = P_0 \,\big|\, G_\tau = P \big]
		\; = \;
		M_\tau(\boldsymbol{\lambda}) [P \to P_0].
	\]
	Substituting this into the previous expression yields exactly \autoref{equ: admixture fraction moments}, completing the proof.
\end{proof}

Furthermore, the time-weighted absorption matrix provides a framework to study the pace of admixture dynamics. The \emph{time to absorption} \(T\) of a gene genealogy \(G\) is defined as the number of generations separating the current generation \(\tau\) from the most recent ancestral configuration in which all alleles reside in source populations:
\[
	T
	\; = \;
	\inf \big\{ t \leq \tau \big| G_{\tau - t} \in \mathcal{S}_2 \big\}.
\]
For instance, in the scenario depicted in \autoref{fig: admixture history}, the time to absorption for the two sampled alleles is \(T = 3\) generations. This quantity reflects the speed at which recent admixture events overwrite older ones, replacing ancient ancestry patterns by more recent gene flows.

\begin{proposition}
	\label{prop: time to absorption moments}
	Let \(T\) denote the time to absorption of the joint gene genealogy \(G\) starting from the transient configuration \(P\) in generation \(\tau\). Then, for any integer \(p \geq 0\), the \(p\)-th moment of \(T\) is given by
	\[
		\mathbf{E} \big[ T^p \big]
		\; = \;
		\sum_{P_0 \in \mathcal{S}_2^{(n)}} M^{(p)}_\tau \big[ P \to P_0 \big].
	\]
\end{proposition}

\begin{proof}
	The time to absorption \(T\) of the gene genealogy \(G\) is defined as the smallest integer such that \(G_{\tau - T}\) is an absorbing configuration. Starting from the transient configuration \(P\), the event \([T = t]\) corresponds to the genealogical paths of the form
	\[
		G_\tau = P
			\quad \to \quad
		\cdots
			\quad \to \quad
		G_{\tau - t + 1} = P'
			\quad \to \quad
		G_{\tau - t} = P_0
			\quad \big(\to \quad
		\cdots
			\quad \to \quad
		G_0 = P_0 \big),
	\]
	with \(P_0 \in \mathcal{S}_2^{(n)}\), and such that \(P' \not\in \mathcal{S}_2^{(n)}\), \textit{i.e.}, such that absorption occurs precisely at generation \(\tau - t\).

	By definition of the time-weighted absorption matrix \(M^{(p)}_\tau\), the entry \(M^{(p)}_\tau \big[ P \to P_0 \big]\) is the weighted sum over all absorbing times:
	\[
		M^{(p)}_\tau \big[ P \to P_0 \big]
		\; = \;
		\sum_{t=0}^{\tau-1} (\tau-t)^p \mathbf{P} \big[ G_t = P_0 \big| G_\tau = P, \; G_{t+1} \not= P_0 \big].
	\]
	Summing over all absorbing configurations \(P_0 \in \mathcal{S}_2^{(n)}\) produces the full moment of \(T\) starting from \(P\):
	\[
		\sum_{P_0 \in \mathcal S_2^{(n)}} M^{(p)}_\tau \big[P \to P_0 \big]
		\; = \;
		\sum_{t=0}^{\tau-1} (\tau-t)^p \mathbf{P} \big[ T = \tau-t \big]
		\; = \;
		\mathbf{E} \big[ T^p \big],
	\]
	completing the proof.
\end{proof}

When a single allele is considered, \(T\) corresponds to the generation of the source genetic ancestor from which it derives. Therefore, \(\mathbf{E} \Big[ T^p \Big| G_0 = \textcolor{vermillion}{\big\{ 1 \big\}_{S_1}} \Big]\) is the \(p\)-th moment of the number of generations \(T\) separating a present-day individual from one of its genetic ancestors randomly sampled from the source population \(\textcolor{vermillion}{S_1}\) \citep{agranat-tamirCountingGeneticAncestors2024}. Adapting the computations in the proof of \autoref{prop: time to absorption moments} yields
\[
	\mathbf{E} \Big[ T^p \Big| G_0 = \textcolor{vermillion}{\big\{ 1 \big\}_{S_1}} \Big]
	\; = \;
	M_\tau^{(p)} \Big[ \big\{ \{ 1 \} \big\} \to \textcolor{vermillion}{\big\{ 1 \big\}_{S_1}} \Big].
\]

\subsection{Stationarity}

Now assume that, after a given generation \(t_0\), population sizes and source contributions become constant, respectively denoted \(N\), \(\textcolor{vermillion}{s_1}\) and \(\textcolor{blue}{s_2}\). This regime, known as \emph{recurrent admixture}, has been studied by \cite{longGeneticStructureAdmixed1991} and \cite{buzbasInferenceAdmixtureFractions2018}, among others. From that time onward, all demographic parameters are fixed, and therefore the transition matrix is also constant; we denote it by \(Q\). Two situations must be distinguished: either the admixed population is isolated (\(\textcolor{vermillion}{s_1} = \textcolor{blue}{s_2} = 0\)), or at least one source population continuously contributes migrants (\(s = \textcolor{vermillion}{s_1} + \textcolor{blue}{s_2} > 0\)). In both cases, a form of stationarity is eventually reached, but its origin differs, and the resulting genetic patterns are not the same. We formalize this in the following theorem and prove it in the two cases separately (\autoref{lem: stationarity in case of isolation} and \autoref{lem: stationarity in case of ongoing gene flow}) to clarify the mechanisms driving stationarity.

\begin{theorem}
	\label{thm: stationarity}
	If, after a given generation \(t_0\), the admixed population has constant size \(N\) and receives constant contributions \(\textcolor{vermillion}{s_1}\) and \(\textcolor{blue}{s_2}\) from the source populations, then, for every \(n \geq 1\) and every \(\boldsymbol{\lambda} \in [0, \Lambda]^n\), the sequence of absorption matrices \(\big( M_\tau(\boldsymbol{\lambda}) \big)_{\tau \geq 0}\) converges to a limiting matrix \(M(\boldsymbol{\lambda})\) as \(\tau \to \infty\).
\end{theorem}

In particular, the admixture fraction reaches a stationary distribution (\autoref{cor: admixture fraction stationarity}), whose structure reflects the post-\(t_0\) regime. Under isolation, the limiting distribution retains a detailed imprint of the admixture process prior to \(t_0\). By contrast, under ongoing gene flow, ancient admixture history is gradually overwritten, and the limiting distribution reflects only the long-term contributions of the source populations.

\begin{corollary}
	\label{cor: admixture fraction stationarity}
	Under the assumptions of \autoref{thm: stationarity}, the admixture fraction \(H^{(\tau)}_{\textcolor{vermillion}{S_1}}\) of a randomly sampled individual in generation \(\tau\), with respect to the source population \(\textcolor{vermillion}{S_1}\), converges in distribution to a limiting random variable \(H_{\textcolor{vermillion}{S_1}}\) as \(\tau \to \infty\).
\end{corollary}

\begin{proof}
	Let \(p\) denote a positive integer. By \autoref{prop: admixture fraction moments},
	\[
		\mathbf{E} \Big[ \big( H^{(\tau)}_{\textcolor{vermillion}{S_1}} \big)^p \Big]
		\; = \;
		\frac{1}{\Lambda^p} \int_{[0, \Lambda]^p} \frac{1}{2^p} \sum_{k = 0}^p \binom{p}{k} M_\tau(\boldsymbol{\lambda}) \Big[ \Big\{ \big\{ \{ 1, \ldots, k \}, \{ k+1, \ldots, p \} \big\} \Big\} \to \Big\{ \textcolor{vermillion}{\big\{ 1, \ldots, p \big\}_{S_1}} \Big\} \Big] \mathrm{d}\boldsymbol{\lambda}.
	\]
	By \autoref{thm: stationarity}, the matrices \(M_\tau(\boldsymbol{\lambda})\) converge pointwise in \(\boldsymbol{\lambda}\) to matrices \(M(\boldsymbol{\lambda})\). Morever, since these matrices are stochastic, all their entries are bounded by \(1\). Hence, applying the dominated convergence theorem yields
	\[
		\mathbf{E} \Big[ \big( H^{(\tau)}_{\textcolor{vermillion}{S_1}} \big)^p \Big]
		\; \longrightarrow \;
		\frac{1}{\Lambda^p} \int_{[0, \Lambda]^p} \frac{1}{2^p} \sum_{k = 0}^p \binom{p}{k} M(\boldsymbol{\lambda}) \Big[ \Big\{ \big\{ \{ 1, \ldots, k \}, \{ k+1, \ldots, p \} \big\} \Big\} \to \Big\{ \textcolor{vermillion}{\big\{ 1, \ldots, p \big\}_{S_1}} \Big\} \Big] \mathrm{d}\boldsymbol{\lambda}.
	\]

	Since the admixture fraction \(H^{(\tau)}_{\textcolor{vermillion}{S_1}}\) takes values in \([0, 1]\), all its moments, as well as their limits, lie within \([0, 1]\). By Theorem 30.1 in \cite{billingsleyProbabilityMeasure1995}, these limiting moments uniquely determine a distribution on \([0, 1]\), denoted \(H_{\textcolor{vermillion}{S_1}}\). By Theorem 30.2 in \cite{billingsleyProbabilityMeasure1995}, the admixture fraction \(H^{(\tau)}_{\textcolor{vermillion}{S_1}}\) converges in distribution to \(H_{\textcolor{vermillion}{S_1}}\) as \(\tau \to \infty\), completing the proof.
\end{proof}

\subsubsection{Isolation}

If the admixed population is isolated, genetic drift progressively homogenizes the population, leading to stationarity. Within the population, alleles eventually reach fixation, so that entire chromosomes become identical across individuals and genetic diversity is lost. In particular, all admixed individuals share the same admixture fraction, which remains a random variable determined by the stochasticity of the early generations.

\begin{lemma}
	\label{lem: stationarity in case of isolation}
	If the admixed population is isolated, \textit{i.e.}, \(\textcolor{vermillion}{s_1} = \textcolor{blue}{s_2} = 0\), then, for every \(n \geq 1\) and every \(\boldsymbol{\lambda} \in [0, \Lambda]^n\), the sequence of absorption matrices \(\big( M_\tau(\boldsymbol{\lambda}) \big)_{\tau \geq 0}\) converges to a rank-one limiting matrix \(M(\boldsymbol{\lambda})\) as \(\tau \to \infty\).
\end{lemma}

\begin{proof}
	Once isolation begins, the transition kernel becomes constant and defines a time-homogeneous Markov chain on the finite configuration space \(\mathcal{H}_N\). This chain is ergodic, and its unique stationary law determines the limit of \(M_\tau(\boldsymbol{\lambda})\).

	\medskip
	\noindent\textbf{Restricted configuration space.}
	When the admixed population is isolated, no parents can be drawn from source populations, so gene genealogies cannot leave the subset \(\mathcal{H}_N \subset \mathcal{P}_N\). For every \(\boldsymbol{\lambda}\), the time-homogeneous transition kernel \(Q(\boldsymbol{\lambda})\) can thus be restricted to \(\mathcal{H}_N\).

	In what follows, we assume for simplicity that the sampled loci \(\boldsymbol{\lambda} = \big( \lambda_1, \lambda_2, \ldots, \lambda_n \big)\) are pairwise distinct. This assumption marks a key distinction from the ongoing gene flow regime. When two or more alleles are located on the same locus, they coalesce with strictly positive probability, and therefore merge within an almost surely finite number of generations. Consequently, within the space of all loci vectors \([0, \Lambda]^n\), subsets where \(\boldsymbol{\lambda}\) contains \(k\) repeated entries exhibit dynamics equivalent to those of the lower-dimensional space \([0, \Lambda]^{n-k+1}\). This effective dimensional collapse induces discontinuous behaviors and eventually breaks the uniformity in the convergence toward stationarity.
	
	\medskip
	\noindent\textbf{Stationarity.}
	Under this assumption, the transition kernel \(Q(\boldsymbol{\lambda})\) is ergodic on \(\mathcal{H}_N\) (details in \autoref{app: ergodicity of the gene genealogies in isolated admixed population}). In other words, coalescence and recombination events mix the sampled alleles within the ancestral population, allowing every allele configuration in \(\mathcal{H}_N\) to be reached. Consequently, there exists a unique stationary distribution \(\mu(\boldsymbol{\lambda})\) such that
	\[
		Q(\boldsymbol{\lambda})^\tau
		\; \longrightarrow \;
		\mathbf{1} \mu(\boldsymbol{\lambda})
		\quad \textrm{as } \tau \to \infty,
	\]
	where \(\mathbf{1}\) denotes the vector of ones.

	Next, consider the matrix
	\[
		\widetilde{R}(\boldsymbol{\lambda})
		\; = \;
		M_{t_0}(\boldsymbol{\lambda})
		\; = \;
		\sum_{t = 0}^{t_0 - 1} \widetilde{Q}_{t_0 - 1}(\boldsymbol{\lambda}) \widetilde{Q}_{t_0 - 2}(\boldsymbol{\lambda}) \cdots \widetilde{Q}_{t + 1}(\boldsymbol{\lambda}) R_t(\boldsymbol{\lambda}),
	\]
	which encodes admixture occuring prior to isolation. Then, for any \(\boldsymbol{\lambda} \in [0, \Lambda]^n\),
	\[
		M_\tau(\boldsymbol{\lambda})
		\; = \;
		\widetilde{Q}(\boldsymbol{\lambda})^{\tau - t_0} \widetilde{R}(\boldsymbol{\lambda})
		\; \longrightarrow \;
		\mathbf{1} \mu(\boldsymbol{\lambda}) \widetilde{R}(\boldsymbol{\lambda})
		\quad \textrm{as } \tau \to \infty,
	\]
	completing the proof.
\end{proof}

Consider the joint gene genealogy \(G\) of two alleles at loci \(\lambda_1\) and \(\lambda_2\), assuming a constant-size admixed population that has been isolated since its foundation. To represent the transition kernels in matrix form, we fix an arbitrary ordering of the configuration space \(\mathcal{P}_N^{(2)}\). First, there are three transient configurations in which both alleles are carried within the admixed population:
\[
	\Big\{ \big\{ \{ 1 \} \big\}, \big\{ \{ 2 \} \big\} \Big\},
	\quad
	\Big\{ \big\{ \{ 1 \}, \{ 2 \} \big\} \Big\},
	\quad
	\Big\{ \big\{ \{ 1, 2 \} \big\} \Big\}.
\]
Second, there are four additional transient configurations corresponding to partial, but incomplete, absorption, in which exactly one marginal genealogy has reached a source population:
\[
	\Big\{ \big\{ \{ 1 \} \big\}, \textcolor{vermillion}{\big\{ \{ 2 \} \big\}_{S_1}} \Big\},
	\quad
	\Big\{ \big\{ \{ 1 \} \big\}, \textcolor{blue}{\big\{ \{ 2 \} \big\}_{S_2}} \Big\},
	\quad
	\Big\{ \textcolor{vermillion}{\big\{ \{ 1 \} \big\}_{S_1}}, \big\{ \{ 2 \} \big\} \Big\},
	\quad
	\Big\{ \textcolor{blue}{\big\{ \{ 1 \} \big\}_{S_2}}, \big\{ \{ 2 \} \big\} \Big\}.
\]
Finally, there are four absorbing configurations, corresponding to complete absorption of both marginal genealogies into the source populations:
\[
	\Big\{ \textcolor{vermillion}{\big\{ 1, 2 \big\}_{S_1}} \Big\},
	\quad
	\Big\{ \textcolor{vermillion}{\big\{ 1 \big\}_{S_1}}, \textcolor{blue}{\big\{ 1 \big\}_{S_2}} \Big\},
	\quad
	\Big\{ \textcolor{vermillion}{\big\{ 2 \big\}_{S_1}}, \textcolor{blue}{\big\{ 1 \big\}_{S_2}} \Big\},
	\quad
	\Big\{ \textcolor{blue}{\big\{ 1, 2 \big\}_{S_2}} \Big\}.
\]
This ordering allows us to refine the block decomposition of the transition kernel \(Q\) (\autoref{equ: transition matrix decomposition}) as
\begin{equation}
	\label{equ: two alleles transition matrix decomposition}
	Q(\boldsymbol{\lambda})
	\; = \;
	\begin{bmatrix}
		\widetilde{Q}_2(\boldsymbol{\lambda}) & \widetilde{Q}_1 & R_2 \\
		\mathbf{0} & h I_4 & R_1 \\
		\mathbf{0} & \mathbf{0} & I_4
	\end{bmatrix},
\end{equation}
where \(I_4\) denotes the \(4 \times 4\) identity matrix. This block structure highlights the hierarchical organization of the dynamics: transitions among fully admixed configurations, transitions involving marginal absorption, and eventual absorption into the source populations.

In this isolated scenario, marginal genealogies cannot be absorbed before foundation. Formally, \(\widetilde{Q}_1 = \mathbf{0}\), \(R_1 = \mathbf{0}\) and \(R_2 = \mathbf{0}\) for \(t > 0\), so that, by \autoref{thm: absorbing markov chain}, the relevant transition kernel reduces to
\[
	\widetilde{Q}_2(\boldsymbol{\lambda})
	\; = \;
	\begin{bmatrix}
		(N-1) \big/ N 	& 1 \big/ 2N 					& 1 \big/ 2N 	\\
		1				& 0								& 0				\\
		0				& \rho(\boldsymbol{\lambda})	& 1 - \rho(\boldsymbol{\lambda})
	\end{bmatrix},
\]
where \(\rho(\boldsymbol{\lambda}) = \rho_{{1}, {2}}^{{1, 2}} \big( \lambda_1, \lambda_2 \big)\) denotes the probability that an odd number of crossovers occur between loci \(\lambda_1\) and \(\lambda_2\), thereby separating the two alleles when they reside on the same chromosome. At foundation, every admixed individual has two parents from source populations, leading to the foundation kernel
\[
	R_2
	\; = \;
	\begin{bmatrix}
		\textcolor{vermillion}{s_1(0)}^2 & \textcolor{vermillion}{s_1(0)} \textcolor{blue}{s_2(0)} & \textcolor{vermillion}{s_1(0)} \textcolor{blue}{s_2(0)} & \textcolor{blue}{s_2(0)}^2 \\
		\textcolor{vermillion}{s_1(0)}^2 & \textcolor{vermillion}{s_1(0)} \textcolor{blue}{s_2(0)} & \textcolor{vermillion}{s_1(0)} \textcolor{blue}{s_2(0)} & \textcolor{blue}{s_2(0)}^2 \\
		\textcolor{vermillion}{s_1(0)} & 0 & 0 & \textcolor{blue}{s_2(0)}
	\end{bmatrix},
\]
while \(\widetilde{Q}_2\) is no longer applicable for \(t = 0\).

Solving a linear system, we obtain the (unique) stationary distribution \(\mu(\boldsymbol{\lambda})\) of \(\widetilde{Q}_2(\boldsymbol{\lambda})\):
\[
	\mu(\boldsymbol{\lambda})
	\; = \;
	\begin{bmatrix}
		\displaystyle
		\frac{2N \rho(\boldsymbol{\lambda})}{2 N \rho(\boldsymbol{\lambda}) + 2\rho(\boldsymbol{\lambda}) + 1}	&
		\displaystyle
		\frac{2\rho(\boldsymbol{\lambda})}{2 N \rho(\boldsymbol{\lambda}) + 2\rho(\boldsymbol{\lambda}) + 1}	&
		\displaystyle
		\frac{1}{2N \rho(\boldsymbol{\lambda}) + 2 \rho(\boldsymbol{\lambda}) + 1}
	\end{bmatrix}.
\]
Accordingly, for any transient configuration \(P \in \mathcal{H}_N\), the stationary absorption probability into the source population \(\textcolor{vermillion}{S_1}\) is
\[
	M(\boldsymbol{\lambda}) \Big[ P \to \Big\{ \textcolor{vermillion}{\big\{ 1, 2 \big\}_{S_1}} \Big\} \Big]
	\; = \;
	\big(\mu(\boldsymbol{\lambda}) R_2\big)_1
	\; = \;
	\frac{2N \rho(\boldsymbol{\lambda}) + 2\rho(\boldsymbol{\lambda})}{2N \rho(\boldsymbol{\lambda}) + 2\rho(\boldsymbol{\lambda}) + 1} \textcolor{vermillion}{s_1(0)}^2 + \frac{1}{2 N \rho(\boldsymbol{\lambda}) + 2\rho(\boldsymbol{\lambda}) + 1} \textcolor{vermillion}{s_1(0)}.
\]

Especially, the first two moments of the stationary admixture fraction follow from \autoref{prop: admixture fraction moments}:
\[
	\mathbf{E}\big[ H_{\textcolor{vermillion}{S_1}} \big]
	\; = \;
	\textcolor{vermillion}{s_1(0)},
	\quad
	\mathbf{E} \big[ H_{\textcolor{vermillion}{S_1}}^2 \big]
	\; = \;
	\frac{1}{\Lambda^2} \int_0^\Lambda \int_0^\Lambda \frac{ 2(N+1) \rho\big(\lambda_1, \lambda_2\big) \textcolor{vermillion}{s_1(0)}^2 + \textcolor{vermillion}{s_1(0)}}{2(N+1) \rho\big(\lambda_1, \lambda_2\big) + 1} \mathrm{d}\lambda_1 \mathrm{d}\lambda_2,
\]
and the stationary variance is
\[
	\mathbf{V} \big[ H_{\textcolor{vermillion}{S_1}} \big]
	\; = \;
	\textcolor{vermillion}{s_1(0)} \big( 1 - \textcolor{vermillion}{s_1(0)} \big)
	\frac{1}{\Lambda^2} \int_0^\Lambda \int_0^\Lambda \frac{\mathrm{d}\lambda_1 \mathrm{d}\lambda_2}{2(N+1) \rho \big(\lambda_1, \lambda_2\big) + 1}.
\]
For example, under the rigid chromosome model (\autoref{equ: rigid recombination probability}), the stationary variance becomes
\begin{equation}
	\label{equ: stationary variance in isolation for rigid chromosomes and Wright-Fisher mating}
	\mathbf{V} \big[ H_{\textcolor{vermillion}{S_1}} \big]
	\; = \;
	\textcolor{vermillion}{s_1(0)} \big( 1 - \textcolor{vermillion}{s_1(0)} \big) \frac{\big(1 + 2(N+1) \rho_0\big) \log \big(1 + 2(N+1) \rho_0\big) - 2(N+1) \rho_0}{2 \big((N+1) \rho_0 \big)^2}.
\end{equation}
All dependence on effective population size \(N\) and recombination rate \(\rho_0\) enters through the factor \((N+1)\rho_0\), reflecting the combined effects of drift and recombination on linkage disequilibrium. Details of all these computations are provided in the \autoref{app: stationary variance in isolation for rigid chromosomes and Wright-Fisher mating}.

In isolation, genetic drift leads to fixation, so the variance within a population is zero. In other words, the stationary variance above captures the variability across possible pedigrees randomness, reflecting our uncertainty about the realized one. When \((N+1)\rho_0 \to \infty\), our model reduces to that of \cite{verduGeneralMechanisticModel2011}, and, as they observed, the variance vanishes. More precisely, in our setting, the remaining variance is of order \(O\big( \log \big( N \rho_0 \big) \big/ N \rho_0 \big)\) (\autoref{fig: law of total variance}, right). Although at most one recombination event can occur on each chromosome per generation, the number of generations effectively recovers their infinite-genome-length hypothesis. Conversely, when \((N+1)\rho_0 \to 0\), the chromosome experiences no recombination, and the stationary variance converges to its maximum value, \(\textcolor{vermillion}{s_1(0)} \big( 1 - \textcolor{vermillion}{s_1(0)} \big)\). This limit coincides with the variance of a Bernoulli random variable, corresponding to a genome whose entire origin is determined by a single draw according to the founding source contributions.

\subsubsection{Ongoing gene flow}

If the admixed population continues to receive migrants, ongoing admixture gradually overwrites past genetic contributions, eventually resulting in stationarity.

\begin{lemma}
	\label{lem: stationarity in case of ongoing gene flow}
	If the admixed population receives an ongoing gene flow from source populations, \textit{i.e.}, \(s = \textcolor{vermillion}{s_1} + \textcolor{blue}{s_2} > 0\), then for every \(n \geq 1\) and \(p \geq 1\), the sequence of absorption matrices \(\big( M_\tau^{(p)}(\boldsymbol{\lambda}) \big)_{\tau \geq 0}\) converges uniformly in \(\boldsymbol{\lambda} \in [0, \Lambda]^n\) to limiting matrices \(M^{(p)}(\boldsymbol{\lambda})\) as \(\tau \to \infty\).
\end{lemma}

\begin{proof}
	Under ongoing gene flow, the contribution of ancient admixture is progressively diluted: the probability that a marginal genealogy remains in a transient configuration decays exponentially fast with time. This exponential control implies that, for each order \(p\), the sequence of matrices \(M_\tau^{(p)}(\boldsymbol{\lambda})\) is Cauchy, and hence converge.

	\medskip
	\noindent\textbf{Absorption probability bound.}
	Consider the joint gene genealogy of \(n\) alleles at loci \(\boldsymbol{\lambda} \in [0, \Lambda]^n\) starting at generation \(\tau\). For each of the \(n\) marginal genealogies, the probability that it is not absorbed before generation \(t\), for any \(t \geq t_0\), is exactly \((1-s)^{\tau-t-1}\). Since absorption of the joint genealogy at generation \(t\) requires that at least one of the marginal genealogies remains unabsorbed until that generation, an union bound yields
	\[
		\sum_{P_0 \in \mathcal{S}_2} \widetilde{Q}_{\tau-1}(\boldsymbol{\lambda}) \widetilde{Q}_{\tau-2}(\boldsymbol{\lambda}) \cdots \widetilde{Q}_{t-1}(\boldsymbol{\lambda}) R_{t}(\boldsymbol{\lambda}) \big[ P \to P_0 \big]
		\; \leq \;
		n (1-s)^{\tau-t-1},
	\]
	independently of the starting and ending configurations \(P\) and \(P_0\), and of loci \(\boldsymbol{\lambda}\). For \(t \leq t_0\), a coarser bound suffices: absorption at such generations requires at least one marginal genealogy to survive up to \(t_0\), yielding
	\[
		\sum_{P_0 \in \mathcal{S}_2^{(n)}} \widetilde{Q}_{\tau-1}(\boldsymbol{\lambda}) \widetilde{Q}_{\tau-2}(\boldsymbol{\lambda}) \cdots \widetilde{Q}_{t-1}(\boldsymbol{\lambda}) R_{t}(\boldsymbol{\lambda}) \big[ P \to P_0 \big]
		\; \leq \;
		n (1-s)^{\tau-t_0-1}.
	\]

	\medskip
	\noindent\textbf{Cauchy property.}
	Let \(\tau_0\) be a positive integer and consider the difference
	\[
		\begin{aligned}
			M_{\tau+\tau_0}^{(p)}(\boldsymbol{\lambda}) - M_\tau^{(p)}(\boldsymbol{\lambda})
			\; = \;
			& \sum_{t = \tau_0}^{\tau + \tau_0 - 1} (\tau + \tau_0 - t)^p \big( \widetilde{Q}_{\tau+\tau_0-1}(\boldsymbol{\lambda}) \cdots \widetilde{Q}_{t+1}(\boldsymbol{\lambda}) R_t(\boldsymbol{\lambda}) - \widetilde{Q}_{\tau-1}(\boldsymbol{\lambda}) \cdots \widetilde{Q}_{t-\tau_0+1}(\boldsymbol{\lambda}) R_{t-\tau_0}(\boldsymbol{\lambda}) \big)
			\\
			& + \sum_{t = 0}^{\tau_0 - 1} (\tau + \tau_0 - t)^p \widetilde{Q}_{\tau+\tau_0-1}(\boldsymbol{\lambda}) \cdots \widetilde{Q}_{t+1}(\boldsymbol{\lambda}) R_t(\boldsymbol{\lambda}).
		\end{aligned}
	\]
	In the first sum, terms cancel whenever \(\tau_0 - t \geq t_0\). Applying the triangle inequality and grouping terms, for any transient configuration \(P\) and absorbed configuration \(P_0\), we have
	\[
		\begin{aligned}
			\Big| M_{\tau+\tau_0}^{(p)}(\boldsymbol{\lambda}) - M_\tau^{(p)}(\boldsymbol{\lambda}) \Big| \big[ P \to P_0 \big]
			\; \leq \;
			& \sum_{t = t_0}^{\tau_0 + t_0 - 1} (\tau + \tau_0 - t)^p \widetilde{Q}_{\tau+\tau_0-1}(\boldsymbol{\lambda}) \cdots \widetilde{Q}_{t+1}(\boldsymbol{\lambda}) R_t(\boldsymbol{\lambda}) \big[ P \to P_0 \big]
			\\
			& + \sum_{t = 0}^{t_0 - 1} (\tau - t)^p \widetilde{Q}_{\tau-1}(\boldsymbol{\lambda}) \cdots \widetilde{Q}_{t+1}(\boldsymbol{\lambda}) R_t(\boldsymbol{\lambda}) \big[ P \to P_0 \big]
			\\
			& + \sum_{t = 0}^{t_0 - 1} (\tau + \tau_0 - t)^p \widetilde{Q}_{\tau+\tau_0-1}(\boldsymbol{\lambda}) \cdots \widetilde{Q}_{t+1}(\boldsymbol{\lambda}) R_t(\boldsymbol{\lambda}) \big[ P \to P_0 \big].
		\end{aligned}
	\]
	Each term corresponds to absorption at generation \(t\). Using the previously established bounds, this gives
	\[
		\Big| M_{\tau+\tau_0}^{(p)}(\boldsymbol{\lambda}) - M_\tau^{(p)}(\boldsymbol{\lambda}) \Big| \big[ P \to P_0 \big]
		\; \leq \;
		(1-s)^{\tau} \Bigg( \sum_{t = t_0}^{\tau_0 + t_0 - 1} \frac{n (\tau + \tau_0 - t)^p}{(1 - s)^{t + 1}} + \sum_{t = 0}^{t_0 - 1} \frac{n (\tau - t)^p}{(1 - s)^{t_0 + 1}} + \sum_{t = 0}^{t_0 - 1} \frac{n (\tau + \tau_0 - t)^p}{(1 - s)^{- \tau_0 + t_0 + 1}} \Bigg).
	\]
	The right-hand side is uniform in configurations \(P\) and \(P_0\), and in loci \(\boldsymbol{\lambda}\), and converges to \(0\) as \(\tau \to \infty\). Therefore, the sequence \(\big( \boldsymbol{\lambda} \mapsto M_\tau(\boldsymbol{\lambda}) \big)\) is Cauchy with respect to the supremum norm of \([0, \Lambda]^n\), and consequently converges uniformly, completing the proof.
\end{proof}

In the case of ongoing gene flow, stationarity is achieved through the convergence of absorption matrices of all orders. As a result, the time to absorption \(T_\tau\) also converges in distribution by \autoref{prop: time to absorption moments}, in contrast to the isolation scenario where \(T_\tau \geq \tau - t_0\) necessarily diverges as \(\tau \to \infty\).

\begin{corollary}
	Under the assumptions of \autoref{lem: stationarity in case of ongoing gene flow}, the time to absorption \(T_\tau\) of a gene genealogy starting in a configuration \(P\) at generation \(\tau\) converges in distribution to a limiting almost surely finite random variable \(T\) as \(\tau \to \infty\).
\end{corollary}

\begin{proof}
	By \autoref{prop: time to absorption moments} together with \autoref{lem: stationarity in case of ongoing gene flow}, all moments of the time to absorption \(T_\tau\) converge to finite limits as \(\tau \to \infty\). Moreover, using the bounds established in the proof of \autoref{lem: stationarity in case of ongoing gene flow}, the \(p\)-th moment of \(T_\tau\) satisfies
	\[
		\mathbf{E} \big[ T_\tau^p \big]
		\; = \;
		\sum_{P_0 \in \mathcal{S}_2} M_\tau^{(p)} \big[ P \to P_0 \big]
		\; \leq \; \sum_{t \geq t_0} (\tau-t)^p (1-s)^{\tau-t-1} s + \sum_{t < t_0} (\tau-t)^p (1-s)^{\tau-t_0-1},
	\]
	where the second term is bounded by
	\[
		O( \tau^p ) (1 - s)^{\tau - t_0 - 1}
		\; \longrightarrow \;
		0
		\quad \textrm{as } \tau \to \infty.
	\]
	It follows that
	\[
		\limsup_{\tau \to \infty} \mathbf{E} \big[ T_\tau^p \big]
		\; \leq \;
		\sum_{t' = 1}^\infty t'^p (1-s)^{t'-1} s
		\; \leq \;
		\frac{p!}{s^p}.
	\]
	This growth bound implies that the associated moment generating power series has radius of convergence at least \(s > 0\). Consequently, Theorem~30.1 in \cite{billingsleyProbabilityMeasure1995} ensures that the limiting moments uniquely determine a probability distribution \(T\). By Theorem~30.2, \(T_\tau\) converges in distribution to \(T\) as \(\tau \to \infty\). Finally, since the limiting expectation is finite, \(T\) is almost surely finite, completing the proof.
\end{proof}

Consider once again the gene genealogy \(G\) of two alleles at loci \(\lambda_1\) and \(\lambda_2\), but this time assuming a constant-size admixed population subject to ongoing gene flow after some generation \(t_0\). For \(\tau \geq t_0\), the absorption matrices satisfy the linear recurrence
\[
	M_{\tau + 1}
	\; = \;
	\widetilde{Q} M_\tau + R,
\]
which implies that the stationary absorption matrix can be expressed as
\[
	M
	\; = \;
	\big( I_7 - \widetilde{Q} \big)^{-1} R,
\]
where \(I\) denotes the \(7 \times 7\) identity matrix. The invertibility of \(I_7 - \widetilde{Q}\) follows from Theorem~11.4 in \cite{grinsteadIntroductionProbability1998}.

Exploiting the refined block decomposition (\autoref{equ: two alleles transition matrix decomposition}), the stationary absorption matrix can be obtained by block inversion:
\[
	M
	\; = \;
	\big( I_7 - \widetilde{Q}(\boldsymbol{\lambda}) \big)^{-1} R
	\; = \;
	\begin{bmatrix}
		\big( I_3 - \widetilde{Q}_2(\boldsymbol{\lambda}) \big)^{-1} \Big( R_2 + \frac{1}{s} \widetilde{Q}_1 R_1 \Big) \\
		\frac{1}{s} R_1
	\end{bmatrix}.
\]
The upper block corresponds to absorption probabilities starting from fully admixed transient configurations, accounting for both direct absorption and absorption following an intermediate marginal event, while the lower block captures absorption from partially absorbed configurations. By \autoref{prop: admixture fraction moments}, the first two moments of the stationary admixture fraction are obtained from the relevant coefficients: \((4, 1)\) for the expected value, corresponding to the dynamics of a single allele, and \((2, 1)\) and \((3, 1)\) for the second moment, reflecting the joint dynamics of two alleles. Therefore, the stationary variance can be computed, as well as higher-order moments in principle.

The computation under the perfectly flexible chromosome model is provided in \autoref{app: stationary variance under ongoing gene flow for perfectly flexible chromosomes and Wright-Fisher mating}. The result is
\begin{equation}
	\label{equ: stationary variance under ongoing gene flow for perfectly flexible chromosomes and Wright-Fisher mating}
	\mathbf{V} \big[ H_{\textcolor{vermillion}{S_1}} \big]
	\; = \;
	\frac{\textcolor{vermillion}{s_1}\textcolor{blue}{s_2}}{s^2} \frac{A}{B} \Bigg( 1 + \frac{1}{\Lambda} \log \Bigg( \frac{B}{C} \Bigg) + \frac{1}{\Lambda^2} \Bigg( \operatorname{Li}_2 \Bigg( \frac{D}{B} \exp(-2\Lambda) \Bigg) - \operatorname{Li}_2 \Bigg( \frac{D}{B} \Bigg) \Bigg) \Bigg),
\end{equation}
where \(\operatorname{Li}_2\) denote the dilogaritm, and
\[
	\begin{aligned}
		A \; = \; N (1-h^2) + h^2,
		& \qquad
		B \; = \; N (1+h) (2-h) + h^2 (2+h),
		\\
		C \; = \; 2 N (1-h^2) + h^2 (2-h^2),
		& \qquad
		D \; = \; h(1+h)(N+h^2).
	\end{aligned}
\]
Under ongoing gene flow, each admixed individual traces its ancestry back to the source populations exponentially fast. Consequently, in sufficiently large populations, admixture fractions behave as independent across individuals within the population. The total variance then reduces to the within-population variance, with no contribution from pedigree randomness.

In the large-population (\(N \to \infty\)) and large-genome (\(\Lambda \to \infty\)) settings, our model again reduces to that of \cite{verduGeneralMechanisticModel2011}, with stationary variance \(\textcolor{vermillion}{s_1}\textcolor{blue}{s_2} \big/ (1+s)s\). Conversely, when the chromosome experiences no recombination (\(\Lambda \to 0\)), the stationary variance converges to \(\textcolor{vermillion}{s_1}\textcolor{blue}{s_2} \big/ 2 s^2 \), corresponding to the variance of the mean of two Bernoulli random variables, representing the origins of two independent homologous chromosomes with relative source contributions \( \textcolor{vermillion}{s_1} \big/ s \) and \( \textcolor{blue}{s_2} \big/ s \). Without taking \(N \to \infty\), genetic drift introduces a slight dependence between chromosome origins, resulting in a small correction to this Bernoulli variance (details in \autoref{app: stationary variance under ongoing gene flow for perfectly flexible chromosomes and Wright-Fisher mating}).

The absorption matrix (\autoref{equ: absorption matrix}) encodes the full distribution of local ancestries conditional on the sampled loci \(\boldsymbol{\lambda}\) and individuals \(P\). Once a mating pattern is specified (\textit{e.g.}, the Wright-Fisher matings), this framework allows derivation of admixture fraction moments (\autoref{prop: admixture fraction moments}), treated as a summary statistic of these local ancestries. However, when multiple alleles are considered, the matrices become high-dimensional and closed-form expressions are generally not computationally feasible. Consequently, only the first moments of the admixture fractions can be explicitly derived, \textit{e.g.}, the variance (\autoref{equ: stationary variance in isolation for rigid chromosomes and Wright-Fisher mating} and \autoref{equ: stationary variance under ongoing gene flow for perfectly flexible chromosomes and Wright-Fisher mating}). Nevertheless, these theoretical results, and particularly the asymptotic analyses, provide a benchmark for understanding how mating patterns shape ancestry patterns arising from admixture history. To further investigate the model’s behavior, we performed simulations across a range of parameter settings, including several simplified extensions designed to capture alternative mating patterns. The simulations illustrate the theoretical predictions and allow exploration of regimes that are otherwise hardly (or not) tractable, thereby connecting model properties to observable patterns in admixed populations.

\subsection{Law of total variance}

\autoref{prop: admixture fraction moments} points out that the variance is the lowest-order moment of the admixture fraction that is sensitive to linkage disequilibrium, that is, to correlations between gene genealogies along the chromosome. In the stationary regimes, accurate approximations can be derived (\autoref{cor: admixture fraction stationarity}). Here, we use simulations to further investigate the behavior of the variance.

\begin{figure*}
	\centering
	\includegraphics{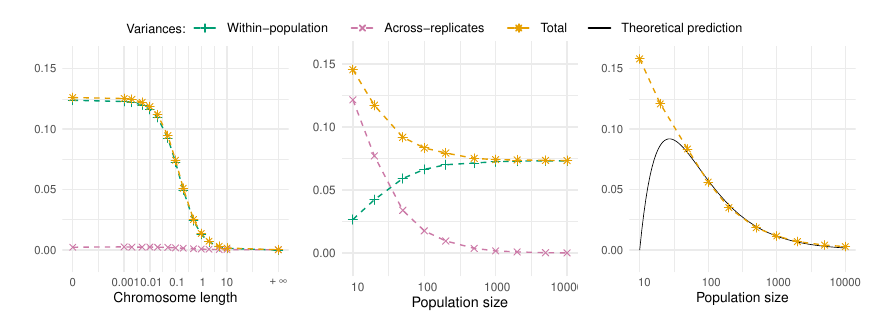}
	\caption{\textbf{The law of total variance decomposes the variance of the admixture fraction into a within-population and an across-replicates components.} \\
	The within-population (green), across-replicates (pink) and total (orange) variances at generation \(\tau = 20\) are shown as functions of chromosome length \(\Lambda\) (left, \(N = 1000\)) and population size \(N\) (center, \(\Lambda = 0.1\)). A zero chromosome length corresponds to a single locus, whereas an infinite length corresponds to a perfectly mixed genome, as in \cite{verduGeneralMechanisticModel2011}. The total variance at stationarity is also shown (right, generation \(\tau\) such that the within-population standard deviation is below \(0.05\)) and compared with the function \(N \mapsto \log(\Lambda N) \big/ 4 \Lambda N\) (black, \(\Lambda = 0.1\)). In all panels, the admixed population is isolated after being founded by two source populations with equal contributions.}
	\label{fig: law of total variance}
\end{figure*}

By the law of total variance,
\[
	\mathbf{V} \big[ H_{\textcolor{vermillion}{S_1}} \big]
	\; = \;
	\mathbf{E} \big[ \mathbf{V} \big[ H_{\textcolor{vermillion}{S_1}} \big| \Pi, \boldsymbol{\Gamma}, \boldsymbol{\gamma} \big] \big]
	+
	\mathbf{V} \big[ \mathbf{E} \big[ H_{\textcolor{vermillion}{S_1}} \big| \Pi, \boldsymbol{\Gamma}, \boldsymbol{\gamma} \big] \big],
\]
where \(\boldsymbol{\Gamma}\) and \(\boldsymbol{\gamma}\) denote the full collection of crossover and segregation events. As noted by \cite{liangUnderstandingAdmixtureFractions2024}, the first term captures the variability \emph{within} an admixed population, while the second reflects variability \emph{across} realizations of the underlying admixture history. In this sense, the second term quantifies the additional variance introduced by implicitly averaging admixture fractions over all possible pedigrees. For instance, in an isolated population of non-recombining chromosomes, the within-population stationary variance of the admixture fraction is zero, as fixation of a single ancestry is eventually reached. However, the total stationary variance (\autoref{equ: stationary variance in isolation for rigid chromosomes and Wright-Fisher mating}) is \(\textcolor{vermillion}{s_1}\big(1-\textcolor{vermillion}{s_1}\big)\), because across replicates the ancestry of the fixed chromosome is determined by a Bernoulli draw.

To illustrate, consider a simple model in which two source populations each contribute half of the admixed population at foundation, after which the population remains isolated. Admixture fractions are recorded at generation \(\tau = 20\). For each parameter setting, we computed the \emph{within-population}, \emph{across-replicates} and total variances using \(1{,}000\) independent replicates, varying either the chromosome length \(\Lambda\) or the (simplified as constant) population size \(N\) (\autoref{fig: law of total variance}). Mating follows the biparental Wright-Fisher model, and recombination is modeled as a Poisson point processes (\autoref{equ: perfectly flexible recombination probability}). This scenario can be interpreted as a schematic representation of the trans-Atlantic slave trade, with African and European source populations.

The within-population variance varies strongly with chromosome length, that is, with the genome mixing rate, but only weakly and indirectly with population size. Any apparent dependence on population size arises through its effect on the coalescence rate, and thus on the speed at which stationarity is reached. If the number of generations \(\tau\) were scaled according to coalescence time, the within-population variance would appear essentially constant with respect to population size. Conversely, the across-replicates variance is primarily sensitive to population size and largely insensitive to chromosome length. In other words, variability among admixed individuals is mainly driven by genetic linkage disequilibrium, as long as the effects of genetic drift leading to fixation remain negligible, whereas variability among admixed populations is driven by \emph{genealogical} linkage disequilibrium, whereby parental relationships can bias the mean admixture fraction before within-population variablity vanishes.

We also simulated the stationary variance for chromosomes of length \(\Lambda = 0.1\) (\autoref{fig: law of total variance}, right), a case in which the perfectly flexible and rigid chromosome models nearly coincide. After waiting for the within-population variance to decay below \(0.0025\), we recorded the total variance. The resulting values closely follow the theoretical rate \(O \big( \log(N \rho_0) \big/ N \rho_0 \big)\) predicted for isolated populations (\autoref{equ: stationary variance in isolation for rigid chromosomes and Wright-Fisher mating}). Together with the other simulations, this indicates that the pedigree induces genealogical linkage disequilibrium, and that its impact on the distribution of admixture fractions is hardly negligible. For instance, quantifying this effect by the approximate across-replicates standard deviation of \(\sqrt{\log(\Lambda N) \big/ 4 \Lambda N}\) for rigid chromosomes under Wright-Fisher matings shows that keeping this deviation below \(0.01\) would require populations larger than \(250,000\) (reproductive) individuals.

\subsection{Demographic scenarios}

The biparental Wright-Fisher model with admixture already exhibits complex behavior as a function of population-level demographic parameters. Introducing the pedigree allows for the construction of even richer demographic scenarios, in which a few extreme events can induce substantial linkage disequilibrium, as observed by \cite{wakeleyEffectsPopulationPedigree2016} and \cite{diamantidisBurstsCoalescencePopulation2024}. We illustrate this flexibility through two models: one for endogamous admixed populations and another incorporating selective pressures. The key idea is that diverse demographic scenarios can be encoded through parameterized distributions for the pedigree \(\Pi\), while preserving the same genetic dynamics. In practice, only a single realization of the admixture history happened, so that only within-population moments can be computed from observed data. Accordingly, we hereafter focus on these moments only.

\begin{figure*}
	\centering
	\includegraphics{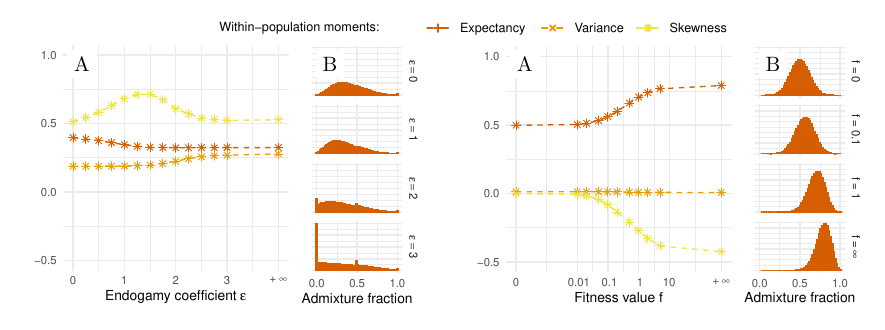}
	\caption{\textbf{Pedigree distributions parameterized to model endogamy (left) and selection (right) alter within-population admixture-fraction moments (A) and distribution (B).} \\
	The within-population moments (A, expectation in red, variance in orange and skewness in yellow) at generation \(\tau = 20\) are shown as functions of the endogamy coefficient \(\varepsilon\) (left, ongoing admixture with contribution \(0.05\)) and the fitness coefficient \(f\) (right, isolated admixed population). Adjacent to the moments plots, selected underlying distributions (B) are displayed as histograms pooling realizations across replicates for each parameter setting. This pooling smooths the distributions without substantially altering them relative to a single realization. In all panels, the admixed population has constant size \(N = 1000\), and individuals carry perfectly flexible chromosomes of length \(\Lambda = 1.0\).}
	\label{fig: demographic scenarios}
\end{figure*}

\subsubsection{Endogamous and exogamous admixed populations}

Human populations often deviate from standard model assumptions, such as the uniformly random mating perscribed by the Wright-Fisher model. Indeed, individuals may choose mates who genetically resemble themselves, a phenomenon known as \emph{assortative mating} \citep{versluysWhyWePick2021}. However, similarities are not restricted to genetic traits. A common cultural practice known as \emph{endogamy} involves choosing mates from within a particular group. Because endogamous marriage can increase the frequency of cousin marriage across different degrees of kinship, we introduce a simple model for endogamy based on time to common ancestors. This models relies critically on the explicit modeling of the pedigree, and therefore differs fundamentally from assortative-mating models \citep{mas-sandovalGenomicFootprintSocial2023,masseyDifferentiatingMechanismOutcome2025}, which rely solely on genetic similarities.

Consider an endogamy coefficient \(\varepsilon\). For two admixed individuals \(i\) and \(j\) at generation \(t\), let \(T_t(i, j)\) denote the number of generations to their most recent common (genealogical) ancestor, with \(T_t(i, j) = \infty\) if no such ancestor exists within the admixed population. We encode endogamous--exogamous mating preferences through a probability measure \(\pi_t^{(i)}(\varepsilon)\), defined for each individual \(i\) in generation \(t\) by a Boltzmann-Gibbs-type distribution:
\[
	\pi_t^{(i)}(\varepsilon): \{ j \}
	\; \mapsto \;
	\frac{1}{Z_t(i)} \exp\Big(- \varepsilon \min \big( T_t(i, j), 5 \big) \Big),
	\quad \textrm{with }
	Z_t(i)
	\; = \;
	\sum_{\substack{j = 1 \\ j \not= i}}^{N_t} \exp\Big(- \varepsilon \min \big( T_t(i, j), 5 \big) \Big).
\]
If the most recent common ancestor predates four generations, individuals are treated as equally related. This prevents the unrealistic consequence, arising from our convention of assigning infinite times to the most recent common ancestor when none exists in the admixed population, that some individuals would be prohibited from mating. Beyond avoiding such artifacts, this rule can be interpreted as reflecting the assumption that, beyond a certain horizon, genealogical relatedness is typically not remembered. Consequently, the distinction between endogamous and exogamous mating becomes less pronounced, as many individuals are effectively undifferentiated in their relatedness.

When both parents are drawn from the admixed population, the first parent \(I_1\) is chosen uniformly from generation \(t\), and then the second parent \(I_2\) is chosen according to \(\pi_t^{(I_1)}(\varepsilon)\). For large positive \(\varepsilon\), the distribution assigns higher probability to close relatives of \(I_1\), modeling endogamy; for small or negative \(\varepsilon\), the abundance of unrelated individuals induces exogamous mating, that is, couples of unrelated parents. When \(\varepsilon = 0\), \(\pi_t^{(I_1)}(\varepsilon)\) reduces to the uniform distribution without selfing, recovering the Wright–Fisher mating rule, which implicitly assumes exogamy.

To study how endogamous populations respond to incoming migrants, we consider an admixed population of size \(N = 1000\) with chromosomes of length \(\Lambda = 1\), initially composed entirely of individuals from source population \(\textcolor{blue}{S_2}\). Over \(\tau = 20\) generations, a constant migrant contribution of \(\textcolor{vermillion}{s_1(t)} = 0.05\) is imposed (\autoref{fig: demographic scenarios}, left). Recall that small positive values of the endogamy coefficient are insufficient to overcome the effect of large population size: with far more unrelated than related individuals, low-\(\varepsilon\) populations remain effectively exogamous. Therefore, we focus on the transition from the Wright-Fisher mating (\(\varepsilon = 0\)) to extreme endogamy (\(\varepsilon = + \infty\)), which appears to be nearly achieved at \(\varepsilon = 3\) for \(N = 1000\).

Under the biparental Wright-Fisher model with admixture, which represents an exogamous mating, genomes mix extensively, yielding higher mean admixture fraction with respect to migrants and lower variance. In contrast, endogamous populations exhibit clustered admixture, with admixture fractions concentrated around chromosomes of a single origin (\(1\), \(0.5\) or \(0\)). In both extremes, skewness is low, but during the transition, the separation of multiple modes temporarily increases skewness, reflecting a heavy tail toward migrant ancestry, until the mode corresponding to full \(\textcolor{blue}{S_2}\) ancestry dominates.

\subsubsection{Dominant selection in admixed populations}

Admixture induces novel gene combinations and phenotypes, thereby creating new opportunities for natural selection and local adaptation to arise, \textit{e.g.}, immunity adaptations in Central Africa \citep{patinDispersalsGeneticAdaptation2017}, or adaptation to hypoxia at high altitude in the Himalayas \citep{huerta-sanchezAltitudeAdaptationTibetans2014}. This process, known as \emph{adaptative admiture}, generates specific genomic signatures \citep{cuadros-espinozaGenomicSignaturesNatural2022}.

Consider a fitness coefficient \(f\) and a locus \(\lambda \in [0, \Lambda]\). When a parent is drawn from the admixed population, the choice is weighted by \(f\): an individual having at least one of its two chromosomes carrying an allele coming from source population \(\textcolor{vermillion}{S_1}\) at locus \(\lambda\) is chosen with weight \(1+f\), and all others with weight \(1\). For \(f > 0\), descendants of \(\textcolor{vermillion}{S_1}\) at locus \(\lambda\) are favored: for example, \(f = 1\) doubles their chance of being sampled as parent. If \(f = \infty\), sampling is restricted to such individuals when available. Conversely, \(f < 0\) represents a disadvantage, and \(f = 0\) recovers the neutral Wright-Fisher mating rule.

We consider an admixed population of size \(N = 1000\) with chromosomes of length \(\Lambda = 1\), initially founded by equal contributions from two source populations \(\textcolor{vermillion}{S_1}\) and \(\textcolor{blue}{S_2}\), and subsequently isolated until generation \(\tau = 20\) (\autoref{fig: demographic scenarios}, right). Under the neutral biparental Wright-Fisher model, the distribution of individual admixture fractions is symmetric. Introducing selection on locus \(\lambda = 0.2\) breaks this symmetry: increasing the fitness coefficient associated with one source population (\(\textcolor{vermillion}{S_1}\)) leads to a systematic increase in the mean admixture fraction derived from that source. However, recombination progressively erodes physical linkage disequilibrium along the genome, so that even arbitrarily strong selection (\textit{e.g.}, \(f = \infty\)) acting at a single locus does not result in uniform ancestry along chromosomes. As a consequence, a nonzero variance in admixture fractions is maintained, together with a negative skewness reflecting the persistence of a tail biased toward the unselected source population (\(\textcolor{blue}{S_2}\)).

\section{Discussion}

We introduced a highly flexible class of probabilistic models for admixture histories, parameterized by a pedigree \(\Pi\). When constrained by an anthropologically motivated parametric probability distribution, this framework enables the explicit study of demographic scenarios (\textit{e.g.}, bottleneck events) and mating patterns (\textit{e.g.}, endogamous behavior). We focused in details on a minimally structured case, that is, the Wright-Fisher mating, establishing structural properties of the resulting gene genealogies and analyzing the induced distributions of admixture fractions. With the exception of closed-form expressions, these results extend to a broad class of demographic histories, provided that the associated gene genealogies are Markov, a property that can always be ensured by enriching the state space beyond allele configuration alone. For example, in our model for endogamous admixed populations, the gene genealogy \(G\) is not Markov, as pedigree layers depend on previous generations. However, augmenting the state of \(G\) with pairwise times to most recent common ancestor restores Markov property, allowing our general results to apply. To illustrate the flexibility of our approach beyond the proposed toy models, note that, although population sizes parameterize the pedigree distribution in all our examples, this is not a requirement. One could, for instance, start with a (monoparental) branching process and then assign an additional parent to each admixed individual (two parents at foundation), possibly drawing these additional parents from the source populations, wherever needed to obtain a random biparental pedigree.

In particular, our theoretical results are fully applicable to sex-structured admixed populations. Biologically, sex constrains the choice of parents, though this effect is negligible, when averaging over pedigrees, if the sex ratio is roughly balanced \citep{crowIntroductionPopulationGenetics2010}. Beyond biology, however, sex carries a social dimension, structuring gendered relationships and, in turn, shaping patterns of admixture, especially when aligned with other social hierarchies such as race \citep{davisWomenRaceClass1981}. For example, historical African European mating often occurred between African women and European men, resulting in higher African ancestry along the X chromosome compared to autosomes \citep{lindElevatedMaleEuropean2006}. However, our results were derived for autosomes and do not directly extend to X chromosomes, where recombination occurs between the mother’s and the father’s mother’s chromosomes. This sex-specific intergenerational recombination implies that X chromosomes carry complementary information about pedigrees \citep{buffaloGenealogicalLookShared2016}, and their sensitivity to mating patterns deserves dedicated investigation.

Regarding source populations, we made two simplifying assumptions. First, they are assumed reproductively isolated, ensuring they act as absorbing states in genealogies and satisfying the requirements of chromosome painting methods, which infer ancestry by comparing present-day admixed and source genomes. In practice, however, source populations are often not fully isolated, \textit{e.g.}, recurrent Eurasian gene flow into coastal African regions \citep{busbyAdmixtureSubSaharanAfrica2016}. Second, they are treated as effectively infinite, so that internal parental relationships do not affect genetic diversity. This assumption may also be violated, particularly when migrants originate from specific subgroups within what seems to be large source populations, \textit{e.g.}, Part~5 in \cite{eltisAtlasTransatlanticSlave2010}.

Under these assumptions, our framework provides a flexible foundation for formulating well-specified demographic histories and studying them via model selection or inference. In most cases, the likelihood (\autoref{equ: absorption matrix}) is expected to be intractable due to the numerous latent variables that are parental relationships and recombination events. However, our simulation procedures are sufficiently fast to make simulation-based inference methods conceivable. Among these, approximate Bayesian computation has already been successfully applied to the study of admixture histories \citep{fortes-limaComplexGeneticAdmixture2021}, and deep learning approaches have been used in a wide range of biological contexts, \textit{e.g.}, phylogenetic inference  \citep{nesterenkoPhyloformerFastAccurate2025}.

The high-dimensionality of genetic data can lead to situations in which too few simulated data lie sufficiently close to the observed data. In population genetics studies, the curse of dimensionality is often mitigated by relying on a limited set of well-chosen summary statistics \citep{beaumontApproximateBayesianComputation2002,csilleryApproximateBayesianComputation2010}. We focused extensively on the admixture fractions, whose within-population moments are sensitive to the pedigree structure. Nevertheless, additional statistics may capture complementary aspects of the pedigree \citep[\textit{e.g.},][for the \(F_4\)-statistic]{ralphEmpiricalApproachDemographic2019}, and can naturally be studied within our framework. For instance, distributions of continuous ancestry tract lengths depend on linkage disequilibrium, and are widely used to estimate admixture timing taking advantage of physical linkage \citep{gravelPopulationGeneticsModels2012,hellenthalGeneticAtlasHuman2014}. These tract distributions are, however, inseparable from genealogical linkage and could therefore be exploited to study mating patterns, as in \cite{mas-sandovalGenomicFootprintSocial2023} on assortative mating. We therefore argue that analyses of ancestry tracts are of primary interest for any statistical study on demographies underlying admixture histories.

Beyond chromosome paintings, alternative metrics can also inform admixture histories. For instance, runs of homozygosity, derived from alleles’ identities by state, provide estimates of relatedness among individuals \citep{pembertonGenomicPatternsHomozygosity2012,ceballosRunsHomozygosityWindows2018}, and are of high relevance for infering the underlying pedigree structure. More generally, ancestral recombination graphs (ARGs) \citep{hudsonPropertiesNeutralAllele1983,griffithsAncestralInferenceSamples1996,wongGeneralEfficientRepresentation2024} encode all information available on marginal gene genealogy trajectories, that is, coalescence and recombination events. An absorption matrix coefficient gives the likelihood of local ancestry obtained by chromosome painting, by summing over all possible trajectories consistent with the initial “sampling” state and the final absorbing state. Inferring an ARG \citep[\textit{e.g.},][]{dengRobustAccurateBayesian2025} of identical-by-state alleles (\textit{e.g.}, haplotypes) provides information about the pedigree by further constraining these trajectories. ARGs could therefore be used to incorporate complementary information (\textit{e.g.}, runs of homozygosity), and notably information derived from mutations, which are not explicitly accounted for in our work, without requiring any additional modeling \citep[see][for a model of mutations on a fixed pedigree]{ralphEmpiricalApproachDemographic2019}.

Conversely, in our framework, the joint gene genealogy of several alleles sampled at the same locus from different admixed individuals is a coalescent. Along a sequence, the one-locus marginal genealogies at each base pair jointly define a gene genealogy equivalent to the ARG of local ancestry, represented as a tree sequence \citep{kelleherEfficientPedigreeRecording2018}. This equivalence is justified in \autoref{app: equivalence with ancestral recombination graphs}. By directly describing ARGs in admixed populations, joint gene genealogies provide a framework for contextualizing ARGs within admixture histories.

\section*{Code availability}

The scripts used in the simulations are available on GitHub: \url{https://github.com/gdoussonlys/admixture_on_a_fixed_pedigree}.

\section*{Funding sources}

This work was supported by the French Government through the France 2030 investment plan, managed by the French National Research Agency (ANR), under grant agreement No. ANR-23-EXMA-0001.

\section*{Acknowledgments}

We thank Alizée Geffroy and Lucas Rey for their valuable assistance with the variance computations. We also are grateful to the INRAE MIGALE bioinformatics facility (MIGALE, INRAE, 2020. Migale bioinformatics Facility, doi: 10.15454/1.5572390655343293E12) for providing computing resources.


\bibliography{Article}
\addcontentsline{toc}{section}{References}

\appendix

\newpage

\part*{Appendices}

\section{Arbitrary number of source populations}
\label{app: arbitrary number of source populations}

For simplicity, we presented our results for admixture histories involving only two source populations \(\textcolor{vermillion}{S_1}\) and \(\textcolor{blue}{S_2}\). In practice, this restriction is easily relaxed, and we briefly outline how our framework extends to an arbitrary number \(m \geq 2\) of source populations.

Consider \(m\) infinitely large source populations \(S_1, S_2, \ldots, S_m\), contributing to generation \(t\) of an admixed population with rates \(\boldsymbol{s}(t) = \big( s_1(t), s_2(t), \ldots, s_m(t) \big)\). At foundation, these contributions sum to one,
\[
	\|\boldsymbol{s}(0)\|_1
	\; = \;
	s_1(0) + s_2(0) + \cdots + s_m(0)
	\; = \;
	1,
\]
while thereafter they sum to at most one, leaving an additional contribution rate \(h(t) = 1-\|\boldsymbol{s}(t)\|_1 \geq 0\) corresponding to the admixed population.

In this setting, the parents of an admixed individual at generation \(t+1\) are drawn in the source population \(S_p\) with probability \(s_p(t)\), and in the admixed population otherwise, that is, with the remaining probability \(h(t)\). All other processes remain unchanged, from mating patterns within the admixed population (\textit{e.g.}, Wright-Fisher, endogamy, selection) to recombination.

Introducing additional source populations leads to additional absorption states in the gene genealogies. The notation for allele configurations (\autoref{equ: allele configuration}) must therefore be extended. Instead of pairs of subsets of \(\llbracket 1, n \rrbracket\) to represent absorbed alleles within a configuration, let \(\mathcal{S}^{(n)}_m\) denote the collection of all (ordered) \(m\)-tuples of subsets of \(\llbracket 1, n \rrbracket\). Accordingly, \(\mathcal{P}^{(n)}_N\) is extended by taking the inverse image of partitions under the canonical mapping from \(\mathcal{S}^{(n)}_m \times \mathcal{H}^{(n)}_N\) to the collection of sets of \(m+2N\) subsets of \(\llbracket 1, n \rrbracket\). Such allele configurations decompose as
\[
	P
	\; = \; 
	\Big\{
		C_{S_1}, \quad
		C_{S_2}, \quad
		\ldots, \quad
		C_{S_m}, \quad
		\big\{  C_{1, 1}, C_{1, 2} \big\}, \quad
		\big\{  C_{2, 1}, C_{2, 2} \big\}, \quad
		\ldots, \quad
		\big\{  C_{N_t, 1}, C_{N_t, 2} \big\}
	\Big\}.
\]
With these extended notations, \autoref{thm: absorbing markov chain}  and all the subsequent results generalize directly.

\section{Proof of \autoref{thm: absorbing markov chain}}
\label{app: absorbing markov chain}

Following the previous section, \autoref{thm: absorbing markov chain} is straightforward to generalize to \(m\) source populations with contributions \(\boldsymbol{s}(t) = \big( s_1(t), s_2(t), \ldots, s_m(t) \big)\) into generation \(t+1\). Indeed, \autoref{equ: transition matrix} simply becomes
\[
	Q_t(\boldsymbol{\lambda}) \big[ P \to P' \big]
	\; = \;
	\Bigg( \prod_{p = 1}^m s_p(t)^{\nu_{t+1}^{(S_p)}} \Bigg) \Bigg( \frac{h(t)}{2 N_t} \Bigg)^{\nu_{t+1}^{(H)}} \Bigg( \frac{N_t}{N_t-1} \Bigg)^{N^{(2, 2)}_{t+1}} \prod_{j = 1}^{\widetilde{N}_t} \Bigg( \big( 2 (N_t - j + 1) \big) \prod_{(i, \delta) \leftarrow j} \rho_{C_{i, \delta}}^{C'_{j, 1}, C'_{j, 2}}(\boldsymbol{\lambda}) \Bigg)
\]
under the biparental Wright-Fisher model with admixture.

\begin{proof}
	We decompose the transition \(P \to P'\) into three independent subtransitions occurring at the population level, the individual level, and the genetic level. The probability of each subtransition is computed separately, and the full transition probability is obtained by multiplying these independent components. In the following, we consider only chromosomes that carry sampled alleles; chromosomes without sampled alleles do not affect the transition probability.
	
	\medskip
	\noindent\textbf{Population subtransition.}
	The probability that, for all \(p\), each of the \(\nu_{t+1}^{(S_p)}\) chromosomes inherited from the source population \(S_p\) coalesces within that population, and that each of the remaining \(\nu_{t+1}^{(H)}\) chromosomes originates from the admixed population is
	\[
		\Bigg( \prod_{p = 1}^m s_p(t)^{\nu_{t+1}^{(S_p)}}\Bigg) h(t)^{\nu_{t+1}^{(H)}}.
	\]
	This factorization follows directly from the biparental Wright-Fisher model with admixture, in which each chromosome is inherited from a population drawn independently, with probabilities given by contributions.

	\medskip
	\noindent\textbf{Individual subtransition.}
	We classify admixed individuals according to the number of chromosome they inherit from admixed parents. Let \(N^{(a, b)}_{t+1}\) denote the number of individuals in generation \(t+1\) with \(a \leq 2\) chromosomes carrying sampled alleles, \(b \leq a\) of which are inherited from admixed parents. These counts are fully determined by the configurations \(P\) and \(P'\). Consequently, the total number of chromosomes inherited from the admixed population is
	\[
		\nu^{(H)}_{t+1}
		\; = \;
		N^{(1,1)}_{t+1} + N^{(2,1)}_{t+1} + 2 N^{(2,2)}_{t+1}.
	\]
	The probability of realizing these parental relationships within the admixed population has already been incorporated in the population subtransition.

	Let \(p^{(1, 1)}\) be the number of distinct admixed parents contributing to the \(N^{(1, 1)}_{t+1}\) first-type individuals. Choosing these parents from the \(N_t\) individuals in generation \(t\) yields \(\binom{N_t}{p^{(1, 1)}}\) possibilities. The parental configuration \(P'\) specifies how the offspring chromosomes are grouped into “sibships”, and there are \(p^{(1, 1)}!\) ways to assign parents to these sibships. Since each offspring independently chooses its admixed parent, the correct assignment occurs with probability \(1 \big/ N_t^{N^{(1, 1)}_{t+1}}\). Altogether, the probability of drawing the right parental relationships from \(P\) to \(P'\) for the \(N^{(1,1)}_{t+1}\) first-type individuals is
	\[
		\Bigg( \frac{1}{N_t} \Bigg)^{N^{(1, 1)}_{t+1}} \binom{N_t}{p^{(1, 1)}} p^{(1, 1)}!
		\; = \;
		\Bigg( \frac{1}{N_t} \Bigg)^{N^{(1, 1)}_{t+1}} \prod_{j = 1}^{p^{(1, 1)}} (N_t - j + 1).
	\]

	For the \(N^{(2, 1)}_{t+1}\) second-type individuals, the reasoning is identical, except that the \(p^{(1, 1)}\) parents already assigned in the previous step must be excluded. Indeed, when offspring chromosomes are grouped by sibships, those associated with some of the \(N^{(1, 1)}_{t+1}\) previous chromosomes have parents that are already determined. Let \(p^{(2, 1)}\) denote the number of additional contributing parents, that is, the number of newly introduced sibships. Then, the probability of drawing the right parental relationships from \(P\) to \(P'\) for the \(N^{(2,1)}_{t+1}\) second-type individuals, knowing the \(p^{(1, 1)}\) previously identified parents, is
	\[
		\Bigg( \frac{1}{N_t} \Bigg)^{N^{(2, 1)}_{t+1}} \binom{N_t - p^{(1, 1)}}{p^{(2, 1)}} p^{(2, 1)}!
		\; = \;
		\Bigg( \frac{1}{N_t} \Bigg)^{N^{(2, 1)}_{t+1}} \prod_{j = 1}^{p^{(2, 1)}} (N_t - p^{(1, 1)} - j + 1).
	\]

	For the \(N^{(2, 2)}_{t+1}\) individuals inheriting two chromosomes (carrying sampled alleles) from two \emph{distinct} admixed parents, let \(p^{(2, 2)}\) denote the number of additional contributing parents (excluding those already counted). Since selfing is impossible, parents for each individual are chosen dependently with probability \(1 \big/ \big( N_t(N_t-1) \big)\) rather than \(1 \big/ N_t^2\). Thus, the probability of drawing the right parental relationships from \(P\) to \(P'\) for the \(N^{(2,2)}_{t+1}\) last-type individuals, knowing all the \(p^{(1, 1)} + p^{(2, 1)}\) previously identified parents, is
	\[
		\Bigg( \frac{1}{N_t (N_t - 1)} \Bigg)^{N^{(2, 2)}_{t+1}} \binom{N_t - p^{(1, 1)} - p^{(2, 1)}}{p^{(2, 2)}} p^{(2, 2)}!
		\; = \;
		\Bigg( \frac{1}{N_t} \Bigg)^{2 N^{(2, 2)}_{t+1}} \Bigg( \frac{N_t}{N_t - 1} \Bigg)^{N^{(2, 2)}_{t+1}} \prod_{j = 1}^{p^{(2, 2)}} (N_t - p^{(1, 1)} - p^{(2, 1)} - j + 1).
	\]

	Multiplying the three products yields the probability of drawing the right parental relationships from \(P\) to \(P'\) for all the admixed individuals:
	\[
		\prod_{j = 1}^{p^{(1, 1)}} (N_t - j + 1) \prod_{j = 1}^{p^{(2, 1)}} (N_t - p^{(1, 1)} - j + 1) \prod_{j = 1}^{p^{(2, 2)}} (N_t - p^{(1, 1)} - p^{(2, 1)} - j + 1)
		\; = \;
		\prod_{j = 1}^{p^{(1, 1)} + p^{(2, 1)} + p^{(2, 2)}} (N_t - j + 1),
	\]
	where \(p^{(1, 1)} + p^{(2, 1)} + p^{(2, 2)} = \widetilde{N}_t\) is the total number of distinct admixed parents contributing chromosomes carrying sampled alleles. The resulting subtransition probability is therefore
	\[
		\Bigg( \frac{1}{N_t} \Bigg)^{\nu_{t+1}^{(H)}} \Bigg( \frac{N_t}{N_t-1} \Bigg)^{N^{(2, 2)}_{t+1}} \prod_{j = 1}^{\widetilde{N}_t} (N_t - j + 1).
	\]

	\medskip
	\noindent\textbf{Recombination subtransition.}
	For each of the \(\nu_{t+1}^{(H)}\) chromosomes inherited from an admixed parent, crossing over groups the carried alleles \(C_{i, \delta}\) on one chromosome from the corresponding parental pair \(\big\{ C'_{j, 1}, C'_{j, 2} \big\}\) with probability
	\[
		\rho_{C_{i, \delta}}^{C'_{j, 1}, C'_{j, 2}}(\boldsymbol{\lambda}),
	\]
	and this chromosome is transmitted with probability \(1 \big/ 2\). Independence across all such events give the overall recombination probability
	\[
		\frac{1}{2^{\nu_{t+1}^{(H)}}} \prod_{(i, \delta) \leftarrow j} \rho_{C_{i, \delta}}^{C'_{j, 1}, C'_{j, 2}}(\boldsymbol{\lambda}),
	\]
	where each parental chromosome pair \(\big\{ C'_{j, 1}, C'_{j, 2} \big\}\) has been assigned an arbitrary ordering, selecting one of the \(2^{\widetilde{N}_t}\) possible global orderings.

	\medskip
	Combining recombination with the previously derived source and admixed subtransitions yields the transition probability stated in \autoref{equ: transition matrix}, completing the proof.
\end{proof}

\section{Ergodicity of the gene genealogies in isolated admixed population}
\label{app: ergodicity of the gene genealogies in isolated admixed population}

In an isolated admixed population of constant size \(N\), the gene genealogy \(G(\boldsymbol{\lambda})\) of \(n\) sampled alleles at loci \(\boldsymbol{\lambda}\) is restricted to \(\mathcal{H}_N\), the set of configurations in which all ancestral alleles remain within the admixed population. We aim to prove that \(G(\boldsymbol{\lambda})\) is ergodic on  \(\mathcal{H}_N\). Throughout the proof, we assume that the entries of \(\boldsymbol{\lambda}\) are pairwise distinct, \textit{i.e.}, \(\lambda_i \not= \lambda_j\) for all \(i \not= j\).

Our proof focuses solely on ergodicity, which suffices to establish the stated result (\autoref{lem: stationarity in case of isolation}). Stronger statements are possible: in particular, one can show that a Doeblin condition holds, representing a lower bound on the mixing rate of alleles. When restricting attention to any compact subset of loci vectors with pairwise distinct entries, this Doeblin condition can be made uniform in \(\boldsymbol{\lambda}\) by minimizing the mixing rate over a finite covering of the compact set, thereby implying uniform convergence. Such refinements lie beyond the scope of this work, since \autoref{thm: stationarity} is used only to characterize and study the stationary distribution of summary statistics.

\medskip
\noindent\textbf{Irreducibility.}
Fix \(\boldsymbol{\lambda}\) and let \(G\) denote the Markov chain induced by \(Q(\boldsymbol{\lambda})\) on \(\mathcal{H}_N\). For \(P, P' \in \mathcal{H}_N\), we show that \(P'\) is reached from \(P\) with a positive probability. First, successive coalescences of lineages within the same individual lead with positive probability from \(P\) to the fully collapsed configuration
\[
	P^\star
	\; = \;
	\big\{ \{ 1, 2, \ldots, n \} \big\},
\]
where all alleles reside on a single chromosome.

Now enumerate the individuals appearing in \(P'\) by labels \(1, 2, \ldots, k\), with \(k \leq N\). Working backward over at most \(N\) generations, at each step \(i\) there is a positive probability that:
\begin{enumerate}
	\item no coalescence occurs among sampled alleles;
	\item recombination separates exactly the alleles destined for individual \(i\) from the others.
\end{enumerate}
Finally, a recombination event can arrange the two homologous chromosomes of each individual to match \(P'\).

Hence, a path of strictly positive probability connects \(P\) to \(P'\), proving that \(G\) is irreductible on \(\mathcal{H}_N\).
	
\medskip
\noindent\textbf{Aperiodicity.}
Fix \(\boldsymbol{\lambda}\) and let \(G\) denote the Markov chain induced by \(Q(\boldsymbol{\lambda})\) on \(\mathcal{H}_N\). For any configuration \(P \in \mathcal{H}_N\), there is a nonzero probability that in one generation:
\begin{enumerate}
	\item no coalescence occurs;
	\item no recombination event affects any chromosome.
\end{enumerate}
Therefore, the period of \(P\) under transition kernel \(Q(\boldsymbol{\lambda})\) is \(1\), proving that \(G\) is aperiodic and, together with irreducibility, ergodic.

\section{Equivalence with ancestral recombination graphs}
\label{app: equivalence with ancestral recombination graphs}

The ancestral recombination graph (ARG), as defined by \cite{wongGeneralEfficientRepresentation2024} as the \emph{genome ARG} (gARG), is a directed acyclic graph (DAG) whose nodes represent haploid genomes and whose time-oriented edges represent genetic ancestral relationships between them. In a gARG, genomes can coalesce (when several edges come from one node) or be split through recombination (when two edges come into one node). Edges corresponding to recombination events are annotated with their crossover points. In the context of admixture, we adapt this definition slightly to include source populations as endpoints of this process. To this end, we associate each source population with an annotated node: an \emph{admixture ARG} (aARG) is a DAG whose nodes represent either admixed haploid genomes or source populations (\textit{e.g.}, \autoref{fig: admixture history}). More specifically, an aARG is obtained from a gARG by identifying all nodes corresponding to haploid genomes from the same source population and merging them into a single node representing that population (\autoref{fig: genome, admixture and extended admixture ancestral recombination graphs}, left).

An aARG records the order of events but not their precise generations, and represents haploid genomes without specifying their pairing into diploid individuals (generation \(t\) corresponds to the gametes produced by generation \(t-1\)). To obtain an ARG representation equivalent to a joint gene genealogy trajectory, we define an \emph{extended admixture ARG} (eaARG), which augments the aARG with intermediate nodes and undirected edges. Directed edges connect successive-generation nodes representing the same haploid genome whenever transmission involves no coalescence, recombination, or absorption, while undirected edges connect homologous haploid genomes within each admixed diploid individual. Contracting uninterrupted transmission edges and removing pairing edges from an eaARG yields a unique aARG (\autoref{fig: genome, admixture and extended admixture ancestral recombination graphs}, right).

\begin{figure*}
	\centering
	\includegraphics{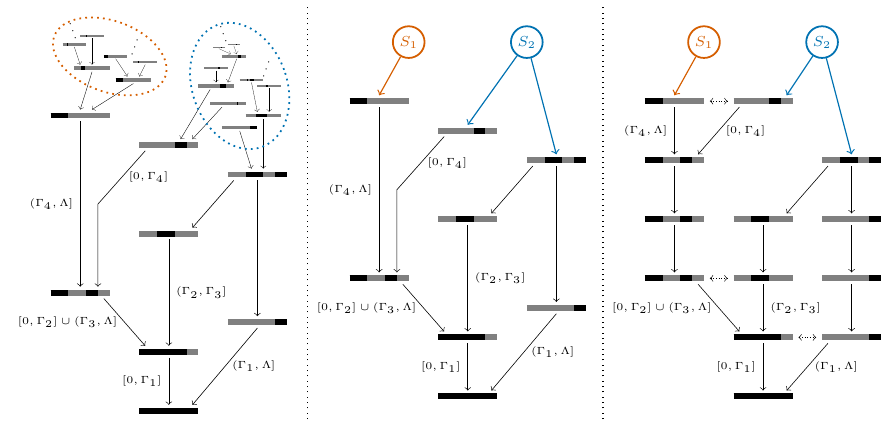}
	\caption{\textbf{Comparison between genome ARG, admixture ARG and extended admixture ARG.}\\
	On the one hand, an aARG (center) is obtained from a gARG (left) by merging all the haploid genomes corresponding to the same source population (dotted ellipses). On the other hand, an aARG (center) is obtained from an eaARG (right) by merging all the uninterrupted transmission edges (one-to-one edges) and discarding the pairing edges (dotted edges). Since source populations are assumed to be infinite, no coalescence occurs within them. Accordingly, the gARG shows only recombination events (splits) deep in the past.}
	\label{fig: genome, admixture and extended admixture ancestral recombination graphs}
\end{figure*}

We first show that any realization of a specific joint gene genealogy uniquely determines an eaARG. Conversely, we show that an eaARG uniquely determines the corresponding joint gene genealogy trajectory. Thus, the two representations are equivalent. In practice, however, only aARGs can be inferred, and the information they retain is insufficient to reconstruct the corresponding eaARG uniquely. This ambiguity reflects irreducible uncertainty in event timing and diploid pairing along the joint gene genealogy trajectory. Although the exact trajectory cannot be recovered, the aARG constrains it to a restricted set of indistinguishable realizations. Event timing matters only through event order, while diploid pairing determines which coalescence and recombination events are possible at each step. Conditional on the aARG, which encodes all events and their order, these unresolved details do not affect the ancestry of the sampled alleles.

\medskip
\noindent\textbf{An eaARG is determined by a joint gene genealogy.}
Consider a sample of partial haploid genomes in generation \(\tau\), represented by chromosome segments \([\Lambda^-_1, \Lambda^+_1], [\Lambda^-_2, \Lambda^+_2], \ldots \subset [0, \Lambda]\). We show how to reconstruct the underlying eaARG from a realization of a specific joint gene genealogy. Note that a small allele sample may fail to capture all absorption, coalescence and recombination events. Biologically, the genealogy of all base pairs in the chromosome segments suffices to identify the segments involved in each coalescence and the locations of recombination events. To formalize this in the continuous-segment setting, consider a countable sample of alleles at loci forming a dense sequence \(\boldsymbol{\lambda} = \big(\lambda_1, \lambda_2, \ldots\big)\) in the disjoint union \([\Lambda^-_1, \Lambda^+_1] \sqcup [\Lambda^-_2, \Lambda^+_2] \sqcup \cdots\) (\textit{e.g.}, rational numbers in these segments), and let \(G\) denote the joint gene genealogy of \(\boldsymbol{\lambda}\).

At generation \(\tau\), the eaARG layer is known: its nodes represent the sampled chromosome segments, and its undirected edges encode their pairing. We reconstruct its earlier layers recursively from the realized trajectory of \(G\). Suppose the eaARG layer at generation \(t\) is known, and let
\[
	G_t
	\; = \; 
	\Big\{
		C_{S_1}, \quad
		C_{S_2}, \quad
		\ldots, \quad
		C_{S_m}, \quad
		\big\{ C_{1, 1}, C_{1, 2} \big\}, \quad
		\big\{ C_{2, 1}, C_{2, 2} \big\}, \quad
		\ldots, \quad
		\big\{ C_{N_t, 1}, C_{N_t, 2} \big\}
	\Big\}
\]
and
\[
	G_{t-1}
	\; = \; 
	\Big\{
		C'_{S_1}, \quad
		C'_{S_2}, \quad
		\ldots, \quad
		C'_{S_m}, \quad
		\big\{ C'_{1, 1}, C'_{1, 2} \big\}, \quad
		\big\{ C'_{2, 1}, C'_{2, 2} \big\}, \quad
		\ldots, \quad
		\big\{ C'_{N_{t-1}, 1}, C'_{N_{t-1}, 2} \big\}
	\Big\},
\]
where \(C_{S_p} \subset C'_{S_p}\) denote the sets of alleles absorbed into source population \(S_p\), and each pair \(\big\{C_{i,1}, C_{i,2}\big\}\) represents the homologous haploid genomes of an admixed diploid individual.

\begin{figure*}
	\centering
	\includegraphics{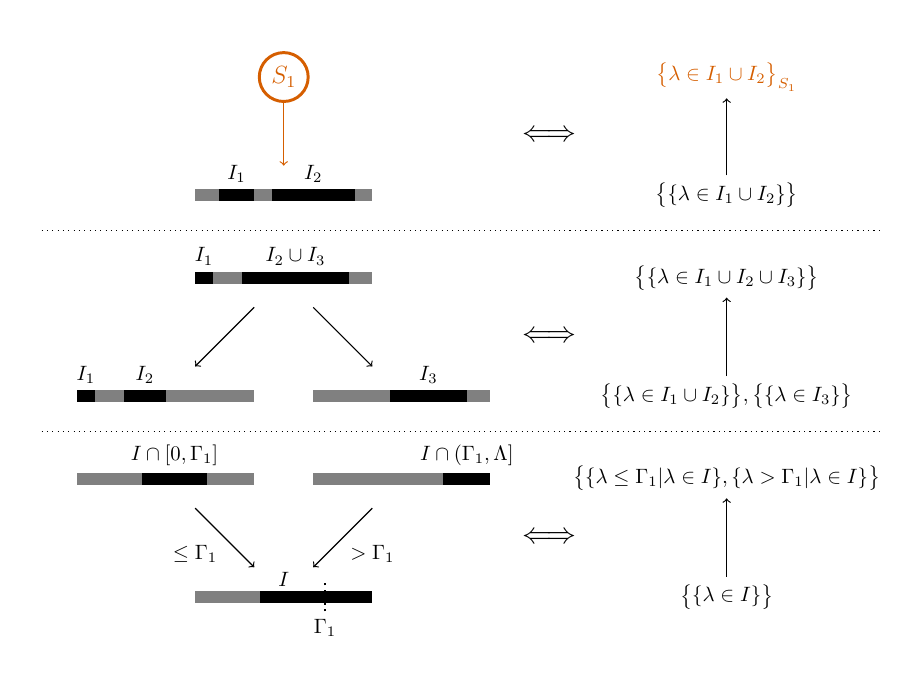}
	\caption{\textbf{All events in eaARGs are encoded in gene genealogies, and reciproquely.}\\
	Absorption (top), coalescence (center) and  recombination (bottom) events are encoded in ARGs, which suffice to describe gene genealogies. Conversely, when sampled alleles are dense along a chromosome segment, these events and their genomic locations are fully determined by the gene genealogy. Consequently, dense gene genealogies and eaARGs are equivalent.}
	\label{fig: joint gene genealogy and ancestral recombination graph equivalence}
\end{figure*}

First, we create one node for each non-empty admixed set \(C'_{i', \delta'}\) in \(G_{t-1}\), and connect the two nodes of each pair \(\big\{C'_{i', 1}, C'_{i', 2}\}\) with an undirected edge. Second, for each pair of sets in \(G_t\) and \(G_{t-1}\) sharing at least one allele, we add a directed edge from the corresponding node at generation \(t-1\) to the node at generation \(t\). If the ancestral set is a source set \(C'_{S_p}\), the edge instead originates from the node representing source population \(S_p\), and represents an absorption event (\autoref{fig: joint gene genealogy and ancestral recombination graph equivalence}, top). Likewise, several edges originating from one node representing an haploid genome corresponds to a coalescence event (\autoref{fig: joint gene genealogy and ancestral recombination graph equivalence}, center).

A recombination event is characterized by one node receiving edges from two distinct nodes: these edges must then be annotated with the crossover points. Without loss of generality, assume that recombination between \(C'_{i', 1}\) and \(C'_{i', 2}\) involved a single crossover point at locus \(\Gamma_1\) and produced the ancestral admixed haploid genome \(C_{i, \delta}\). If the crossover occurs within a contiguous segment ancestral to the sample in generation \(t\), then \(C_{i, \delta}\) contains a neighbourhood of \(\Gamma_1\), \textit{i.e.}, there exists \(\varepsilon > 0\) such that
\[
	\big\{\Gamma_1 - \varepsilon < \lambda_i < \Gamma_1 + \varepsilon \big| i = 1, 2, \dots \big\}
	\;\subset\;
	C_{i, \delta}.
\]
Backward in time, this neighbourhood is split in \(G_{t-1}\), that is,
\[
	\big\{\Gamma_1 - \varepsilon < \lambda_i < \Gamma_1 \big| i = 1, 2, \dots\big\}
	\; \subset \;
	C'_{i', 1}
	\quad \textrm{and} \quad
	\big\{\Gamma_1 \leq \lambda_i < \Gamma_1 + \varepsilon \big| i = 1, 2, \dots\big\}
	\; \subset \;
	C'_{i', 2}.
\]
Because the sequence \(\boldsymbol{\lambda}\) is dense, this split uniquely identifies the crossover point \(\Gamma_1\), even if no allele is sampled at that precise locus (\autoref{fig: joint gene genealogy and ancestral recombination graph equivalence}, right). By contrast, if the crossover happened between two disjoint ancestral segments \([\Lambda_1, \Lambda_2]\) and \([\Lambda_3, \Lambda_4]\), with \(\Lambda_2 < \Lambda_3\), its precise location cannot be recovered. This ambiguity is immaterial, since any crossover point (strictly) between \(\Lambda_2\) and \(\Lambda_3\) yields an equivalent eaARGs.

By recursively applying this construction, the trajectory of \(G\) uniquely determines the underlying eaARG, and hence the aARG, up to the locations of crossovers occurring outside the ancestral segments represented in the sample.

\medskip
\noindent\textbf{Properties of gene genealogies generalize naturally to countably infinite sets of alleles.}
We have shown that an aARG is fully determined by the joint gene genealogy \(G\) of alleles at loci \(\boldsymbol{\lambda}\) dense in all sampled chromosome segments. However, such a genealogy requires a countably infinite set of alleles, rather than the finite set considered previously. We briefly explain why our results extend to this setting.

The genealogy \(G\) remains an absorbing Markov chain, fully absorbed at foundation at the latest. The configuration space can be constructed analogously to the finite case (\textit{e.g.}, as in \autoref{app: arbitrary number of source populations}), replacing the finite index set \(\llbracket 1, n \rrbracket\) with the natural numbers \(\mathbf{N} = \{1, 2, \ldots\}\). Moreover, the combinatoric proof of \autoref{thm: absorbing markov chain} remains valid, as it relies on counting chromosomes, which are still in finite numbers. Similarly all other proofs generalize without modification, except for that of \autoref{lem: stationarity in case of isolation}.

We say that an allele is \emph{unseparated} from the sample if every neighbourhood of its locus contains at least one other sampled allele. Consider an unseparated allele undergoing a recombination event. The point process \(\Gamma\) almost surely contains finitely many crossover points. We additionaly assume that the allele almost surely does not lie on a crossover point, \textit{e.g.}, because the intensity measure of \(\Gamma\) is diffuse (\textit{i.e.}, non-atomic). Thus, there exists a neighbourhood of the allele that contains no crossover point, but contains other sampled alleles by unseparation, implying that the allele remains unseparated from the sample. In other words, transitions of the genealogy \(G\) preserve the property of being unseparated. This subtle feature complicates the proof of \autoref{lem: stationarity in case of isolation}, since genomes can no longer become perfectly mixed in a finite time. Nevertheless, genetic drift eventually leads to fixation, and a stationary state is reached. The lemma therefore remains valid.

In summary, considering a countably infinite set of alleles in the joint gene genealogy \(G\) changes little compared with the finite case.

\medskip
\noindent\textbf{A joint gene genealogy is determined by an eaARG.}
Consider the eaARG containing the chromosome segments \([\Lambda^-_1, \Lambda^+_1], [\Lambda^-_2, \Lambda^+_2], \ldots \subset [0, \Lambda]\) in generation \(\tau\). Let \(\boldsymbol{\lambda}\) denote a finite or countably infinite sequence of loci in the disjoint union \([\Lambda^-_1, \Lambda^+_1] \sqcup [\Lambda^-_2, \Lambda^+_2] \sqcup \cdots\), and let \(G\) denote the joint gene genealogy of the alleles sampled at these loci. We show how to reconstruct the trajectory of \(G\). In particular, when \(\boldsymbol{\lambda}\) is dense in the disjoint union, this construction yields the reciprocal implication, and thus establishes the equivalence of the two representations.

Suppose that the configuration
\[
	G_t
	\; = \; 
	\Big\{
		C_{S_1}, \quad
		C_{S_2}, \quad
		\ldots, \quad
		C_{S_m}, \quad
		\big\{ C_{1, 1}, C_{1, 2} \big\}, \quad
		\big\{ C_{2, 1}, C_{2, 2} \big\}, \quad
		\ldots, \quad
		\big\{ C_{N_t, 1}, C_{N_t, 2} \big\}
	\Big\}
\]
is known, and that every non-empty set \(C_{i, \delta}\) corresponds to a single identified node in the eaARG. Initially, this condition holds only for \(t = \tau\). We now construct the unique possible \(G_{t-1}\), conditional on \(G_t\) and on the eaARG, thereby recursively determining the entire trajectory.

To begin, we split all sets associated with recombination events. For every node corresponding to a set \(C_{i, \delta}\) with two incoming edges, we partition \(C_{i, \delta}\) into two sets, \(C_{i, \delta, 1}\) and \(C_{i, \delta, 2}\), each corresponding to one edge, using the crossover points annotating the edges. For instance, given two crossover points \(\mu_1\) and \(\mu_2\), we would define the two intermediate sets as
\[
	C_{i, \delta, 1}
	\; = \;
	\big\{\lambda < \mu_1 \textrm{ or } \mu_2 \geq \lambda \big| \lambda \in C_{i, \delta}\big\}
	\quad \textrm{and} \quad
	C_{i, \delta, 2}
	\; = \;
	\big\{\mu_1 \geq \lambda < \mu_2 \big| \lambda \in C_{i, \delta}\big\}.
\]
Note that, at this stage, we already know that \(C_{i, \delta, 1}\) and \(C_{i, \delta, 2}\) corresponds to two homologous admixed chromosomes, even without the undirected edge in the eaARG, because recombination occurs only between the two chromosomes of a same diploid individual. We emphasize that the additional information encoded in the eaARG compared with the aARG is of limited importance, as it serves only to distinguish trajectories of \(G\) that are consistent with all events and their ordering.

Then, we only need to follow the edges, allowing for possible merges, to construct the configuration \(G_{t-1}\). Without loss of generality, assume that, following our previous treatment of recombination events, all sets \(C_{i, \delta}\) in \(G_t\) correspond to nodes with only one incoming edge. If the incoming edge of the node corresponding to \(C_{i, \delta}\) originates from the node associated with source population \(S_p\), we include \(C_{i, \delta}\) in \(C'_{S_p}\), which also contains \(C_{S_p}\). If the incoming edge originates from another node, we include \(C_{i, \delta}\) in \(C'_{i', \delta'}\), where \(i'\) and \(\delta'\) identify the edge’s startpoint. Specifically, \(C'_{i', \delta}\) is the union of all sets corresponding to endpoints of edges originating from that node, and \(C'_{i', 1}\) and \(C'_{i', 2}\) are connected with an undirected edge. Thereby, completing with possible empty sets, we uniquely determine the configuration
\[
	G_{t-1}
	\; = \; 
	\Big\{
		C'_{S_1}, \quad
		C'_{S_2}, \quad
		\ldots, \quad
		C'_{S_m}, \quad
		\big\{ C'_{1, 1}, C'_{1, 2} \big\}, \quad
		\big\{ C'_{2, 1}, C'_{2, 2} \big\}, \quad
		\ldots, \quad
		\big\{ C'_{N_{t-1}, 1}, C'_{N_{t-1}, 2} \big\}
	\Big\},
\]
along with the correspondance between sets in \(G_{t-1}\) and nodes in the eaARG.

When the sampled alleles are dense in the sampled haploid genomes, this recursive construction reverses the one described above, and thus recovers the trajectory that generated the eaARG. In other words, a dense joint gene genealogy is equivalent to an eaARG (\autoref{fig: joint gene genealogy and ancestral recombination graph equivalence}, bottom), and is (almost) equivalent to an aARG.

\medskip
\noindent\textbf{An example.}
To illustrate, consider the eaARG shown on the \autoref{fig: genome, admixture and extended admixture ancestral recombination graphs} (right), and construct the equivalent joint gene genealogy. Let \(\boldsymbol{\lambda}\) be a dense sequence of loci spanning the sampled chromosome \([0, \Lambda]\) (bottom). Reading the eaARG yields the full trajectory of its joint gene genealogy \(G(\boldsymbol{\lambda})\):
\[
	\begin{aligned}
		G_6 \; = \; \big\{ \{0 \leq \lambda \leq \Lambda\} \big\}
		\quad \to \quad &
		G_5 \; = \; \big\{ \{0 \leq \lambda \leq \Gamma_1\}, \; \{\Gamma_1 < \lambda \leq \Lambda\} \big\}
		\\
		\quad \to \quad &
		G_4 \; = \; \big\{ \{0 \leq \lambda \leq \Gamma_2 \textrm{ ou } \Gamma_3 < \lambda \leq \Gamma_1\}, \; \{\Gamma_2 < \lambda \leq \Gamma_3\} \big\}, \quad\big\{ \{\Gamma_1 < \lambda \leq \Lambda\} \big\}
		\\
		\quad \to \quad &
		G_3 \; = \; \big\{ \{0 \leq \lambda \leq \Gamma_2 \textrm{ ou } \Gamma_3 < \lambda \leq \Gamma_1\} \big\}, \quad \big\{ \{\Gamma_2 < \lambda \leq \Gamma_3\} \big\}, \quad \big\{ \{\Gamma_1 < \lambda \leq \Lambda\} \big\}
		\\
		\quad \to \quad &
		G_2 \; = \; \big\{ \{0 \leq \lambda \leq \Gamma_2 \textrm{ ou } \Gamma_3 < \lambda \leq \Gamma_1\} \big\}, \quad \big\{ \{\Gamma_2 < \lambda \leq \Gamma_3 \textrm{ ou } \Gamma_1 < \lambda \leq \Lambda\} \big\}
		\\
		\quad \to \quad &
		G_1 \; = \; \big\{ \{0 \leq \lambda \leq \Gamma_2 \}, \; \{\Gamma_3 < \lambda \leq \Gamma_1\} \big\}, \quad \textcolor{blue}{ \big\{ \Gamma_2 < \lambda \leq \Gamma_3 \textrm{ ou } \Gamma_1 < \lambda \leq \Lambda \big\}_{S_2} }
		\\
		\quad \to \quad &
		G_0 \; = \; \textcolor{vermillion}{ \big\{ 0 \leq \lambda \leq \Gamma_2 \big\}_{S_1} }, \quad \textcolor{blue}{ \big\{ \Gamma_2 < \lambda \leq \Lambda \big\}_{S_2} }.
	\end{aligned}
\]
The aARG alone does not determine this trajectory because it records the ordering of events but not their timing. For example, without the eaARG, the relative timing of the independent recombination and coalescence events in the left and right lineages cannot be established.

Because \(\boldsymbol{\lambda}\) is dense in \([0, \Lambda]\), the black intervals at each generation \(t\) are fully characterized by \(G_t\), and hence the eaARG is effectively equivalent to its corresponding gene genealogy \(G\). Every crossover point within the sampled interval is identifiable from the dense sequence, except for \(\Gamma_4\), which lies outside the interval but still affects the sample’s history in the transition \(G_2 \to G_1\).

By contrast, chromosome painting alone reveals only 
the initial configuration \(G_6\), namely the sampling in the admixed population, and the corresponding origins \(G_0\) in terms of source populations. Many trajectories, and thus many distinct eaARGs (and aARGs), are consistent with these observations. Among the crossover points shown, only \(\Gamma_2\) is ultimately visible in the painted chromosome. Therefore, inferring an ARG from identical-by-states alleles provide additional information about demographic history that alleles’ origin alone cannot recover.

\medskip
\noindent\textbf{Towards ARG likelihood?}
Up to few minor additional information, gene genealogies (of countably infinitely many alleles) represent the same object as ARGs in the context of admixture. The properties we derived therefore translate to the ARG framework. In particular, we have obtained a matrix, the absorption matrix \(M\) (\autoref{equ: absorption matrix}), that gives the likelihood of the observed chromosome painting. This matrix is defined by summing over all possible trajectories, but the likelihood of a specific trajectory (or a highly constrained set of trajectories) can be computed directly from the transition probabilities in \(Q\) (\autoref{thm: absorbing markov chain}). Hence, one can compute the likelihood of an ARG in our model.

However, inferring a single underlying ARG is unlikely to be feasible, and we argue that a Bayesian framework is more appropriate (\textit{e.g.}, \citep{dengRobustAccurateBayesian2025}). Moreover, the need to define an extended ARG reveals the numerous trajectories compatible with the ordered events described by a given ARG. Consequently, likelihood-based inference faces the same issue as when using chromosome painting: the computation is intractable. Appendices \ref{app: stationary variance in isolation for rigid chromosomes and Wright-Fisher mating} and \ref{app: stationary variance under ongoing gene flow for perfectly flexible chromosomes and Wright-Fisher mating} illustrate how difficult it is to compute the absorption matrix, even in the case of only two alleles.

\section{Stationary variance of the admixture fraction in isolation}
\label{app: stationary variance in isolation for rigid chromosomes and Wright-Fisher mating}

Consider a population which has \(N\) reproductive individuals in each generation, and is isolated since foundation. For a chromosome segment of small length \(\Lambda\) morgans, the rigid model applies (\autoref{equ: rigid recombination probability}). By \autoref{cor: admixture fraction stationarity}, the admixture fraction of a randomly sampled individual in the generation \(\tau\) converges as \(\tau \to \infty\), and so does its variance. Although the rigid model limits the number of recombination events to one per generation, the many generations separating source ancestors from sampled individuals can compensate for this limitation, so that even long stationary chromosomes may be well described by the following computation.

The stationary variance of the admixture fraction can be computed using \autoref{prop: admixture fraction moments}. In Section 3.2.1, we showed that, during population isolation, the gene genealogy remains within \(\mathcal{H}_N\). The corresponding restricted transition kernel is
\[
	\widetilde{Q}_2(\boldsymbol{\lambda})
	\; = \;
	\begin{bmatrix}
		(N-1) \big/ N 	& 1 \big/ 2N 					& 1 \big/ 2N 	\\
		1				& 0								& 0				\\
		0				& \rho(\boldsymbol{\lambda})	& 1 - \rho(\boldsymbol{\lambda})
	\end{bmatrix},
\]
where \(\rho(\boldsymbol{\lambda}) = \rho_{\{1\}, \{2\}}^{\{1, 2\}} \big(\lambda_1, \lambda_2 \big)\) denotes the probability that an odd number of crossovers occur between loci \(\lambda_1\) and \(\lambda_2\), thereby separating the two alleles when they reside on the same chromosome. Upon reaching the foundation, the gene genealogy is immediately absorbed through
\[
	R_2
	\; = \;
	\begin{bmatrix}
		\textcolor{vermillion}{s_1(0)}^2 & \textcolor{vermillion}{s_1(0)} \textcolor{blue}{s_2(0)} & \textcolor{vermillion}{s_1(0)} \textcolor{blue}{s_2(0)} & \textcolor{blue}{s_2(0)^2} \\
		\textcolor{vermillion}{s_1(0)}^2 & \textcolor{vermillion}{s_1(0)} \textcolor{blue}{s_2(0)} & \textcolor{vermillion}{s_1(0)} \textcolor{blue}{s_2(0)} & \textcolor{blue}{s_2(0)^2} \\
		\textcolor{vermillion}{s_1(0)} & 0 & 0 & \textcolor{blue}{s_2(0)}
	\end{bmatrix},
\]
since all founding parents come from source populations.

\noindent\textbf{Stationary distribution.} According to \autoref{lem: stationarity in case of isolation}, we first need to compute the stationary distribution \(\mu\) of \(\widetilde{Q}_2(\boldsymbol{\lambda})\) to obtain \(M\), by solving
\[
	\mu \widetilde{Q}_2(\boldsymbol{\lambda})
	\; = \;
	\mu,
\]
which gives
\[
	\begin{cases}
		\mu_1 \frac{N-1}{N} + \mu_2
		& \; = \;
		\mu_1
		\\
		\mu_1 \frac{1}{2N} \phantom{+// \mu_2} + \mu_3 \rho(\boldsymbol{\lambda})
		& \; = \;
		\mu_2
		\\
		\mu_1 \frac{1}{2N} \phantom{+// \mu_2} + \mu_3 \big(1-\rho(\boldsymbol{\lambda})\big)
		& \; = \;
		\mu_3,
	\end{cases}
\]
with the additional conditions \(\mu_1 \geq 0\), \(\mu_2 \geq 0\), \(\mu_3 \geq 0\), and \(\mu_1 + \mu_2 + \mu_3 = 1\) since \(\mu\) is a probability distribution.

The unique solution of this system is
\[
	\mu
	\; = \;
	\begin{bmatrix}
		\displaystyle
		\frac{2N \rho(\boldsymbol{\lambda})}{2N \rho(\boldsymbol{\lambda}) + 2 \rho(\boldsymbol{\lambda}) + 1}	&
		\displaystyle
		\frac{2 \rho(\boldsymbol{\lambda})}{2N \rho(\boldsymbol{\lambda}) + 2 \rho(\boldsymbol{\lambda}) + 1}	&
		\displaystyle
		\frac{1}{2N \rho(\boldsymbol{\lambda}) + 2 \rho(\boldsymbol{\lambda}) + 1}.
	\end{bmatrix}.
\]
Indeed,
\[
	\begin{cases}
		 \big(2N \rho(\boldsymbol{\lambda})\big) \frac{N-1}{N} + \big(2\rho(\boldsymbol{\lambda})\big)
		& \; = \;
		2N \rho(\boldsymbol{\lambda})
		\\
		\big(2N \rho(\boldsymbol{\lambda})\big) \frac{1}{2N} + \big(1\big) \rho(\boldsymbol{\lambda})
		& \; = \;
		2\rho(\boldsymbol{\lambda})
		\\
		\big(2N \rho(\boldsymbol{\lambda})\big) \frac{1}{2N} + \big(1\big) \big(1-\rho(\boldsymbol{\lambda})\big)
		& \; = \;
		1
	\end{cases}
\]

\noindent\textbf{Stationary variance.} Due to ergodicity of the isolated gene genealogy, for any configuration \(P\), the stationary absorption probability into the source population \(\textcolor{vermillion}{S_1}\) is
\[
	\begin{aligned}
		M(\boldsymbol{\lambda}) \Big[ P \to \Big\{ \textcolor{vermillion}{\big\{ 1, 2 \big\}_{S_1}} \Big\} \Big]
		& \; = \;
		\mu R_2[\cdot, 1]
		\\
		& \; = \;
		\frac{
		\begin{bmatrix}
			2 N \rho(\boldsymbol{\lambda})
			&
			2 \rho(\boldsymbol{\lambda})
			&
			1
		\end{bmatrix}
		}{{2 N \rho(\boldsymbol{\lambda}) + 2\rho(\boldsymbol{\lambda}) + 1}}
		\begin{bmatrix}
			\textcolor{vermillion}{s_1(0)}^2 \\
			\textcolor{vermillion}{s_1(0)}^2 \\
			\textcolor{vermillion}{s_1(0)}
		\end{bmatrix}
		\\
		& \; = \;
		\frac{(2N+1) \rho(\boldsymbol{\lambda})}{(2N+1) \rho(\boldsymbol{\lambda}) + 1} \textcolor{vermillion}{s_1(0)}^2 + \frac{1}{(2N+1) \rho(\boldsymbol{\lambda}) + 1} \textcolor{vermillion}{s_1(0)},
	\end{aligned}
\]
where \([\cdot, 1]\) denotes extraction of the first column of the preceding matrix. Therefore, by \autoref{prop: admixture fraction moments},
\[
	\begin{aligned}
		\mathbf{E} \big[ H_{\textcolor{vermillion}{S_1}}^2 \big]
		& \; = \;
		\frac{1}{\Lambda^2} \int_0^\Lambda \int_0^\Lambda \frac{M(\boldsymbol{\lambda}) \big[ \big\{\{1, 2\}\big\} \to \big\{ \textcolor{vermillion}{\{1, 2\}_{S_1}} \big\} \big] + M(\boldsymbol{\lambda}) \big[ \big\{\{1\}, \{2\}\big\} \to \big\{ \textcolor{vermillion}{\{1, 2\}_{S_1}} \big\} \big]}{2} \mathrm{d}\lambda_1 \mathrm{d}\lambda_2
		\\
		& \; = \;
		\frac{1}{\Lambda^2} \int_0^\Lambda \int_0^\Lambda \frac{2(N+1) \rho\big( \lambda_1, \lambda_2 \big) \textcolor{vermillion}{s_1(0)}^2 + \textcolor{vermillion}{s_1(0)}}{2 (N+1) \rho\big( \lambda_1, \lambda_2 \big) + 1} \mathrm{d}\lambda_1 \mathrm{d}\lambda_2.
	\end{aligned}.
\]
To obtain the variance, we subtract the squared expectation. The expected admixture fraction is simply the contribution of the source population \(\textcolor{vermillion}{S_1}\) at foundation, \(\textcolor{vermillion}{s_1(0)}\), since a single allele traces back to a source population according to the founding contributions. Thus,
\[
	\begin{aligned}
		\mathbf{V} \big[ H_{\textcolor{vermillion}{S_1}} \big]
		& \; = \;
		\frac{1}{\Lambda^2} \int_0^\Lambda \int_0^\Lambda \frac{ 2(N+1) \rho \big( \lambda_1, \lambda_2 \big) \textcolor{vermillion}{s_1(0)}^2 + \textcolor{vermillion}{s_1(0)}}{2(N+1) \rho \big( \lambda_1, \lambda_2 \big) + 1} \mathrm{d}\lambda_1 \mathrm{d}\lambda_2 - \textcolor{vermillion}{s_1(0)}^2
		\\
		& \; = \;
		\frac{1}{\Lambda^2} \int_0^\Lambda \int_0^\Lambda \frac{2(N+1) \rho \big( \lambda_1, \lambda_2 \big) \textcolor{vermillion}{s_1(0)}^2 + \textcolor{vermillion}{s_1(0)} - \big(2(N+1) \rho \big( \lambda_1, \lambda_2 \big) + 1\big) \textcolor{vermillion}{s_1(0)}^2}{2(N+1) \rho \big( \lambda_1, \lambda_2 \big) + 1} \mathrm{d}\lambda_1 \mathrm{d}\lambda_2
		\\
		& \; = \;
		\textcolor{vermillion}{s_1(0)} \big( 1 - \textcolor{vermillion}{s_1(0)} \big)
		\frac{1}{\Lambda^2} \int_0^\Lambda \int_0^\Lambda \frac{\mathrm{d}\lambda_1 \mathrm{d}\lambda_2}{2(N+1) \rho \big( \lambda_1, \lambda_2 \big) + 1}.
	\end{aligned}
\]

\noindent\textbf{Rigid chromosome model.} To proceed, we need to specify \(\rho\). Since our chromosome segment is small, we consider the simplest realistic model, namely the rigid chromosome model (\autoref{equ: rigid recombination probability}):
\[
	\rho_{\{1\}, \{2\}}^{\{1, 2\}} \big( \lambda_1, \lambda_2 \big)
	\; = \;
	\rho_0 \frac{\big| \lambda_2 - \lambda_1 \big|}{\Lambda}.
\]
This gives
\[
	\begin{aligned}
		\frac{1}{\Lambda^2} \int_0^\Lambda \int_0^\Lambda \frac{\mathrm{d}\lambda_1 \mathrm{d}\lambda_2}{1 + 2(N+1) \rho \big( \lambda_1, \lambda_2 \big)}
		& \; = \;
		\frac{1}{\Lambda^2} \int_0^\Lambda \int_0^\Lambda \frac{\mathrm{d}\lambda_1 \mathrm{d}\lambda_2}{1 + 2(N+1) \rho_0 \frac{|\lambda_2-\lambda_1|}{\Lambda}}
		\\
		& \; = \;
		\int_0^1 \frac{2(1-x)}{1 + 2(N+1) \rho_0 x} \mathrm{d}x.
	\end{aligned}
\]
A probabilistic justification for this change of variables is given in the previous section. Denoting \(a = 2(N+1) \rho_0\), a partial fraction decomposition yields
\[
	\begin{aligned}
		\int_0^1 \frac{2(1-x)}{1+ax} \mathrm{d}x
		& \; = \;
		2 \int_0^1 \frac{a (1-x)}{a (1+ax)} \mathrm{d}x
		\\
		& \; = \;
		2 \int_0^1 \frac{a + 1 - 1 - ax}{a (1+ax)} \mathrm{d}x
		\\
		& \; = \;
		2 \int_0^1  \frac{a + 1}{a (1+ax)} - \frac{1}{a} \mathrm{d}x
		\\
		& \; = \;
		2 \int_0^1  \frac{1 + \frac{1}{a}}{1+ax} \mathrm{d}x - 2 \int_0^1 \frac{1}{a} \mathrm{d}x
		\\
		& \; = \;
		2 \Bigg(1 + \frac{1}{a}\Bigg) \Bigg[\frac{1}{a} \log(1 + ax)\Bigg]_0^1 - \frac{1}{a}
		\\
		& \; = \;
		\frac{2(a+1)\log(1+a) - a}{a^2}.
	\end{aligned}
\]

Finally, the variance can be written as
\[
	\begin{aligned}
		\mathbf{V} \big[ H_{\textcolor{vermillion}{S_1}} \big]
		& \; = \;
		\textcolor{vermillion}{s_1(0)} \big( 1 - \textcolor{vermillion}{s_1(0)} \big) \frac{2(1+a)\log(1+a) - a}{a^2}
		\\
		& \; = \;
		\textcolor{vermillion}{s_1(0)} \big( 1 - \textcolor{vermillion}{s_1(0)} \big) \frac{\big( 1 + 2(N+1) \rho_0 \big) \log \big(1 + 2(N+1) \rho_0\big) - 2(N+1) \rho_0}{2 \big((N+1) \rho_0 \big)^2}.
	\end{aligned}
\]
All dependence on population size \(N\) and recombination rate \(\rho_0\) is made through the factor \((N+1)\rho_0\), so that linkage disequilibrium is due to drift as well as recombination.

\noindent\textbf{Limiting cases.} Denoting \(I = (N+1)\rho_0\), any linkage disequilibrium in a two-genes genealogy is due to the factor \(I\). Indeed,
\[
	\mathbf{V} \big[ H_{\textcolor{vermillion}{S_1}} \big]
	\; = \;
	\textcolor{vermillion}{s_1(0)} \big( 1 - \textcolor{vermillion}{s_1(0)} \big) \frac{(1 + 2I) \log(1 + 2I) - 2I}{2I^2}.
\]

In the model of \cite{verduGeneralMechanisticModel2011}, the admixture fraction in an isolated population converges to \(0\) as \(\tau \to \infty\), due to the fixation of a genome over generations. In our case, their assumption of a large population and long genomes are better, though not perfectly, captured by \(I = (N+1)\rho_0 \to \infty\). In this limit,
\[
	\mathbf{V} \big[ H_{\textcolor{vermillion}{S_1}} \big]
	\; \underset{I \to \infty}{\sim} \;
	\textcolor{vermillion}{s_1(0)} \big( 1 - \textcolor{vermillion}{s_1(0)} \big) \frac{\log I}{I}
	\; \underset{I \to \infty}{\to} \;
	0.
\]
Although the rigid chromosome model we used prevents us from assuming the same perfect mixing by recombination as in \cite{verduGeneralMechanisticModel2011}, the stationary variances coincides: because the recombination rate \(\rho_0\) exceeds than the coalescence rate \(1 \big/ N\), numerous generations elapse before fixation, compensating for the limited number of recombination events per generation, and ultimately yielding the same perfectly mixed and fixed genome. The across-replicates variance vanishes due to perfect mixing, while the within-population variance vanishes due to fixation.

Let \(f\) denote the “linkage function”
\[
	f(I)
	\; = \;
	\frac{(1 + 2I) \log(1 + 2I) - 2I}{2I^2}.
\]
Its derivative is
\[
	f'(I)
	\; = \;
	- \frac{(1+I) \log(1+2I) - 2I}{I^3}.
\]
Defining the numerator in the expression of \(f'\) as \(g: I \mapsto (1+I) \log(1+2I) - 2I\), its two first derivatives are
\[
	g'(I)
	\; = \;
	\log(1+2I) - \frac{2I}{1+2I},
	\textrm{ and }
	g''(I)
	\; = \;
	\frac{4I}{(1+2I)^2}.
\]
Recall that \(I > 0\). Clearly, \(g''(I) > 0\). Since \(g'(0) = 0\), it follows that \(g'(I) > 0\). Moreover, \(g(0) = 0\), so \(g(I) > 0\), and hence \(f'(I) < 0\). Therefore \(f\) is strictly decreasing over \(I > 0\), with maximum value \(1\) reached in the limit \(I = (N+1)\rho_0 \to 0\), corresponding to the absence of recombination. In this case, the stationary variance is maximized:
\[
	\lim_{I \to 0} \mathbf{V} \big[ H_{\textcolor{vermillion}{S_1}} \big]
	\; = \;
	\textcolor{vermillion}{s_1(0)} \big( 1 - \textcolor{vermillion}{s_1(0)} \big).
\]
Under isolation, fixation is always reached at stationarity, so the within-population variance vanishes. The remaining variance is entirely across replicates, and corresponds to choosing a single ancestry for each replicate, with probability given by its source contribution. Thus, the stationary variance reduces to that of a Bernoulli random variable.

To summarize, the linkage disequilibrium is characterized by \(I = (N+1)\rho_0\) through \(f(I)\). As \(I \to \infty\), the stationary variance agrees with that of \cite{verduGeneralMechanisticModel2011}, where the linkage is neglected. As \(I \to 0\), the stationary variance is maximized, corresponding to the strongest possible linkage. Thus, \(I = (N+1)\rho_0\) can be interpreted as a measure of the “degree of independence” between two gene genealogies sampled from the same individual (\autoref{prop: admixture fraction moments}).

\section{Stationary variance of the admixture fraction under ongoing gene flow}
\label{app: stationary variance under ongoing gene flow for perfectly flexible chromosomes and Wright-Fisher mating}

Consider a population which, after a generation \(t_0\), has \(N\) reproductive individuals in each generation, and receives ongoing gene flow at positive rate \(s = \textcolor{vermillion}{s_1} + \textcolor{blue}{s_2} > 0\). Let us focus on a chromosome of length \(\Lambda\) morgans. By \autoref{cor: admixture fraction stationarity}, the admixture fraction of a randomly sampled individual in the generation \(\tau\) converges as \(\tau \to \infty\). In particular, its variance converges, and we aim to compute this limiting variance.

This quantity can be computed using \autoref{prop: admixture fraction moments}. In Section 3.2.2, we showed that the stationary absorption matrix can be written as
\[
	M
	\; = \;
	\big( I_7 - \widetilde{Q}(\boldsymbol{\lambda}) \big)^{-1} R
	\; = \;
	\begin{bmatrix}
		\big( I_3 - \widetilde{Q}_2(\boldsymbol{\lambda}) \big)^{-1} \Big( R_2 + \frac{1}{1-h} \widetilde{Q}_1 R_1 \Big) \\
		\frac{1}{1-h} R_1
	\end{bmatrix}.
\]
By \autoref{thm: absorbing markov chain}, the blocks governing transitions among transient configurations are given by
\[
	\widetilde{Q}_2(\boldsymbol{\lambda})
	\; = \;
	\begin{bmatrix}
		h^2 (N-1) \big/ N 	& h^2 \big/ 2N 								& h^2 \big/ 2N \\
		h^2					& 0											& 0 \\
		0					& h \rho(\boldsymbol{\lambda})	& h \big( 1 - \rho(\boldsymbol{\lambda}) \big)
	\end{bmatrix},
	\quad
	\widetilde{Q}_1
	\; = \;
	\begin{bmatrix}
		h \textcolor{vermillion}{s_1} & h \textcolor{blue}{s_2} & h \textcolor{vermillion}{s_1} & h \textcolor{blue}{s_2} \\
		h \textcolor{vermillion}{s_1} & h \textcolor{blue}{s_2} & h \textcolor{vermillion}{s_1} & h \textcolor{blue}{s_2} \\
		0 & 0 & 0 & 0
	\end{bmatrix},
\]
where \(\rho(\boldsymbol{\lambda}) = \rho_{\{1\}, \{2\}}^{\{1, 2\}} \big(\lambda_1, \lambda_2 \big)\) denotes the probability that an odd number of crossovers occur between loci \(\lambda_1\) and \(\lambda_2\), thereby separating the two alleles when they reside on the same chromosome. The blocks describing transitions into absorbing configurations are
\[
	R_2
	\; = \;
	\begin{bmatrix}
		\textcolor{vermillion}{s_1^2} & \textcolor{vermillion}{s_1} \textcolor{blue}{s_2} & \textcolor{vermillion}{s_1} \textcolor{blue}{s_2} & \textcolor{blue}{s_2^2} \\
		\textcolor{vermillion}{s_1^2} & \textcolor{vermillion}{s_1} \textcolor{blue}{s_2} & \textcolor{vermillion}{s_1} \textcolor{blue}{s_2} & \textcolor{blue}{s_2^2} \\
		\textcolor{vermillion}{s_1} & 0 & 0 & \textcolor{blue}{s_2}
	\end{bmatrix},
	\quad
	R_1
	\; = \;
	\begin{bmatrix}
		\textcolor{vermillion}{s_1} & 0	& \textcolor{blue}{s_2} & 0 \\
		0 & \textcolor{vermillion}{s_1} & 0 & \textcolor{blue}{s_2} \\
		\textcolor{vermillion}{s_1} & 0	& \textcolor{blue}{s_2} & 0 \\
		0 & \textcolor{vermillion}{s_1} & 0 & \textcolor{blue}{s_2} \\
	\end{bmatrix}.
\]

\noindent\textbf{Stationary expectation.} Using the ordering described in Section 3.2.1, we only need to compute the coordinate \((4, 1)\) to obtain the expectation. Indeed, by \autoref{prop: admixture fraction moments},
\[
	\mathbf{E}\big[H_{\textcolor{vermillion}{S_1}}\big]
	\; = \;
	\frac{1}{\Lambda} \int_0^\Lambda M[4, 1](\lambda) \textrm{d}\lambda
	\; = \;
	\frac{1}{\Lambda} \int_0^\Lambda \frac{1}{s} R_1[1, 1] \textrm{d}\lambda
	\; = \;
	\frac{\textcolor{vermillion}{s_1}}{s}.
\]
In other words, the average admixture fraction equals the relative contribution of the source population.

\noindent\textbf{Matrix coefficients.} Similar to the expectation case, we only need to evaluate the \((2, 1)\) and \((3, 1)\) coefficients to compute the stationary variance. Let \(\boldsymbol{m}\) denote the first column of the upper block of \(M\), so that
\[
	\begin{aligned}
		\big(I_3 - \widetilde{Q}_2\big) \boldsymbol{m}
		& \; = \;
		\Bigg( R_2 + \frac{1}{1-h} \widetilde{Q}_1 R_1 \Bigg) [\cdot, 1]
		\\
		& \; = \;
		\begin{bmatrix}
			\textcolor{vermillion}{s_1}^2
			\\
			\textcolor{vermillion}{s_1}^2
			\\
			\textcolor{vermillion}{s_1}
		\end{bmatrix}
		+ 
		\frac{1}{1+h}
		\begin{bmatrix}
			h \textcolor{vermillion}{s_1} & h \textcolor{blue}{s_2} & h \textcolor{vermillion}{s_1} & h \textcolor{blue}{s_2} \\
			h \textcolor{vermillion}{s_1} & h \textcolor{blue}{s_2} & h \textcolor{vermillion}{s_1} & h \textcolor{blue}{s_2} \\
			0 & 0 & 0 & 0
		\end{bmatrix}
		\begin{bmatrix}
			\textcolor{vermillion}{s_1} \\
			0 \\
			\textcolor{vermillion}{s_1} \\
			0
		\end{bmatrix}
		\\
		& \; = \;
		\begin{bmatrix}
			\frac{1+h}{1-h} \textcolor{vermillion}{s_1}^2 &
			\frac{1+h}{1-h} \textcolor{vermillion}{s_1}^2 &
			\textcolor{vermillion}{s_1}
		\end{bmatrix}^T,
	\end{aligned}
\]
where \([\cdot, 1]\) denotes extraction of the first column of the preceding matrix. Cramer’s rule yields
\[
	M[2, 1]
	\; = \;
	\boldsymbol{m}_2
	\; = \;
	\frac{\det K_2}{\det K},
	\textrm{ and }
	M[3, 1]
	\; = \;
	\boldsymbol{m}_3
	\; = \;
	\frac{\det K_3}{\det K},
\]
where \(K\) denote the matrix \(I_3 - \widetilde{Q}_2\), and \(K_j\) the matrix \(K\) where the \(j\)-th column has been replaced by \(\big( R_2 + \frac{1}{1-h} \widetilde{Q}_1 R_1 \big) [\cdot, 1]\). The rule of Sarrus, for instance, gives
\[
	\begin{aligned}
		\displaystyle \det K
		& \; = \;
		\Bigg(1 - h^2\frac{N-1}{N}\Bigg) \big(1 - h \big(1 - \rho(\boldsymbol{\lambda})\big)\big) - \frac{h^2}{2N} h^2 h \rho(\boldsymbol{\lambda}) - \frac{h^2}{2N} h^2 \big(1 - h \big(1 - \rho(\boldsymbol{\lambda})\big)\big)
		\\
		& \; = \;
		\Bigg(1 - h^2 + \frac{h^2}{N}\Bigg) \big(1 - h + h \rho(\boldsymbol{\lambda})\big) - \frac{h^4}{2N} \big(1 - h + 2 h \rho(\boldsymbol{\lambda})\big),
		\\
		\displaystyle \det K_2
		& \; = \;
		\Bigg(1 - h^2 \frac{N-1}{N}\Bigg) \Bigg(\frac{1+h}{1-h} \textcolor{vermillion}{s_1}^2\Bigg) \big(1 - h \big(1 - \rho(\boldsymbol{\lambda})\big)\big) + \frac{h^2}{2N} h^2 \textcolor{vermillion}{s_1} + \Bigg(\frac{1+h}{1-h} \textcolor{vermillion}{s_1}^2\Bigg) h^2 \big(1 - h \big(1 - \rho(\boldsymbol{\lambda})\big)\big) 
		\\
		& \; = \;
		\Bigg(1 + \frac{h^2}{N}\Bigg) \big(1 - h + h\rho(\boldsymbol{\lambda})\big) \frac{1+h}{1-h} \textcolor{vermillion}{s_1}^2 + \frac{h^4}{2N} \textcolor{vermillion}{s_1},
		\\
		\displaystyle \det K_3
		& \; = \;
		\Bigg(1 - h^2 \frac{N-1}{N}\Bigg) \textcolor{vermillion}{s_1} + \Bigg(\frac{1+h}{1-h} \textcolor{vermillion}{s_1}^2\Bigg) h^2 \big(h \rho(\boldsymbol{\lambda})\big) + \Bigg(1 - h^2 \frac{N-1}{N}\Bigg) \Bigg(\frac{1+h}{1-h} \textcolor{vermillion}{s_1}^2\Bigg) \big(h \rho(\boldsymbol{\lambda})\big)  - \frac{h^2}{2N} h^2 \textcolor{vermillion}{s_1}
		\\
		& \; = \;
		\Bigg(1 + \frac{h^2}{N}\Bigg) \big(h \rho(\boldsymbol{\lambda})\big) \frac{1+h}{1-h} \textcolor{vermillion}{s_1}^2 + \Bigg(1 - h^2 + \frac{h^2}{N} - \frac{h^4}{2N}\Bigg) \textcolor{vermillion}{s_1}.
	\end{aligned}
\]

\noindent\textbf{Integral decomposition.} By \autoref{prop: admixture fraction moments}, the second moment is
\[
	\begin{aligned}
		\mathbf{E}\big[H_{\textcolor{vermillion}{S_1}}^2\big]
		& = \;
		\frac{1}{2\Lambda^2} \int_0^\Lambda \int_0^\Lambda M[2, 1] + M[3, 1] \textrm{d}\lambda_1 \textrm{d}\lambda_2
		\\
		& = \;
		\frac{1}{2\Lambda^2} \int_0^\Lambda \int_0^\Lambda \frac{\det K_2 + \det K_3}{\det K} \textrm{d}\lambda_1 \textrm{d}\lambda_2
		\\
		& = \;
		\frac{1}{2\Lambda^2} \int_0^\Lambda \int_0^\Lambda \frac{\Big(1 + \frac{h^2}{N}\Big) \big(1 - h + 2h\rho(\lambda_1, \lambda_2)\big) \frac{1+h}{1-h} \textcolor{vermillion}{s_1}^2 + \Big(1 - h^2 + \frac{h^2}{N}\Big) \textcolor{vermillion}{s_1}}{\Big(1 - h^2 + \frac{h^2}{N}\Big) \big(1 - h + h \rho(\lambda_1, \lambda_2)\big) - \frac{h^4}{2N} \big(1 - h + 2 h \rho(\lambda_1, \lambda_2)\big)} \textrm{d}\lambda_1 \textrm{d}\lambda_2.
	\end{aligned}
\]
The integrand can be written as the rational fraction
\[
	f
	\; = \;
	\frac{a + b \rho}{c + d \rho},
\]
where
\[
	\begin{aligned}
		a
		\; = \;
		\Bigg(1 + \frac{h^2}{N}\Bigg) (1+h) \textcolor{vermillion}{s_1}^2 + \Bigg(1 - h^2 + \frac{h^2}{N}\Bigg) \textcolor{vermillion}{s_1},
		& \qquad
		b
		\; = \;
		2h \Bigg(1 + \frac{h^2}{N}\Bigg) \Bigg(\frac{1+h}{1-h}\Bigg) \textcolor{vermillion}{s_1}^2,
		\\
		c
		\; = \;
		(1-h) \Bigg(1 - h^2 + \frac{h^2}{N} - \frac{h^4}{2N} \Bigg),
		& \qquad
		d
		\; = \;
		h \Bigg(1 - h^2 + \frac{h^2}{N} - \frac{h^4}{N} \Bigg).
	\end{aligned}
\]

The dependence on \(\lambda_1\) and \(\lambda_2\) is entirely through \(\rho\), so a partial fraction expansion separates the integrand in simpler terms:
\[
	\begin{aligned}
		f
		& \; = \;
		\frac{a + b\rho}{c+d\rho}
		\\
		& \; = \;
		\frac{ad - bc + bc + bd\rho}{d (c + d\rho)}
		\\
		& \; = \;
		\frac{ad - bc}{d} \frac{1}{c + d\rho} + \frac{b}{d}.
	\end{aligned}
\]
Consequently,
\[
	\mathbf{E}\big[H_{\textcolor{vermillion}{S_1}}^2\big]
	\; = \;
	\frac{b}{2d} + \frac{ad - bc}{d} \frac{1}{2\Lambda^2} \int_0^\Lambda \int_0^\Lambda \frac{1}{c+d\rho(\lambda_1, \lambda_2)} \textrm{d}\lambda_1 \textrm{d}\lambda_2.
\]

\noindent\textbf{Perfectly flexible chromosome model.} To proceed, we need to specify a model for \(\rho\). Unlike the isolation case, where the rigidity of chromosomes is compensated by the large number of generations, allowing chromosomes to become effectively fully mixed between the foundation and the sampled generation, we must here consider large chromosome segment to compare limiting behaviors with simpler models such as the one of \cite{verduGeneralMechanisticModel2011}. We therefore assume the perfectly flexible chromosome model, that is, \autoref{equ: perfectly flexible recombination probability}:
\[
	\rho\big(\lambda_1, \lambda_2\big)
	\; = \;
	\frac{1 - \exp\big(-2 \big| \lambda_2 - \lambda_1 \big| \big)}{2}.
\]

Under this model, the recombination probability \(\rho\big(\lambda_1, \lambda_2\big)\) between two loci \(\lambda_1\) and \(\lambda_2\) only depends on their distance \(\big| \lambda_2 - \lambda_1 \big|\). Let \(U_1 = \lambda_1 \big/ \Lambda\) and \(U_2 = \lambda_2 \big/ \Lambda\) denote two independent uniform variables on \([0, 1]\), so that
\[
	\frac{1}{2\Lambda^2} \int_0^\Lambda \int_0^\Lambda \frac{1}{c+d\rho\big(\lambda_1, \lambda_2\big)} \textrm{d}\lambda_1 \textrm{d}\lambda_2
	\; = \;
	\frac{1}{2} \mathbf{E}\Bigg[\frac{1}{c+d\rho\big(\Lambda U_1, \Lambda U_2\big)}\Bigg].
\]
Now let \(X = \big|U_1-U_2\big|\). Its cumulative distribution function is
\[
	\mathbf{P}[X \leq x] = 1 - \mathbf{P}\big[\big|U_1-U_2\big| > x\big] = 1 - (1 - x)^2,
\]
because \(\big(U_1, U_2\big)\) is distributed uniformly on the square while the event \(\big\{\big|U_1-U_2\big| > x\big\}\) consists of two triangles, each of side length \(1-x\). Therefore, the density of \(X\) is \(f_X(x) = 2(1-x)\), and
\[
	\frac{1}{2} \mathbf{E}\Bigg[\frac{1}{c+d\rho\big(\Lambda X\big)}\Bigg]
	\; = \;
	\int_0^1 \frac{1-x}{c+d\rho(\Lambda x)} \textrm{d}x,
\]
where \(\rho(x)\) is shorthand for \(\rho\big(\lambda_1, \lambda_2\big)\) when \(x = \big|\lambda_2 - \lambda_1\big|\).

\noindent\textbf{Integration by parts.} \autoref{equ: perfectly flexible recombination probability} gives
\[
	\begin{aligned}
		\int_0^1 \frac{1-x}{c+d\rho(\Lambda x)} \textrm{d}x
		& \; = \;
		\frac{2}{2c+d} \int_0^1 \frac{1-x}{1-\frac{d}{2c+d} \exp(-2 \Lambda x)}
		\\
		& \; = \;
		\frac{2}{2c+d} \int_0^1 (1-x) g(x) \textrm{d}x,
		& \textrm{where } g: x
		\; \longmapsto \;
		\frac{1}{1-\frac{d}{2c+d} \exp(-2 \Lambda x)}.
	\end{aligned}
\]
Note that an antiderivative of \(g\) is
\[
	G: x
	\; \longmapsto \;
	x + \frac{1}{2\Lambda} \log\Bigg(1 - \frac{d}{2c+d} \exp(-2\Lambda x)\Bigg).
\]
Integrating by parts,
\[
	\begin{aligned}
		\int_0^1 (1-x) g(x) \textrm{d}x
		& \; = \;
		-G(0) - \int_0^1 -G(x) \textrm{d}x
		\\
		& \; = \;
		-\frac{1}{2\Lambda} \log\Bigg(1 - \frac{d}{2c+d}\Bigg) + \int_0^1 x + \frac{1}{2\Lambda} \log\Bigg(1 - \frac{d}{2c+d} \exp(-2\Lambda x)\Bigg) \textrm{d}x
		\\
		& \; = \;
		-\frac{1}{2\Lambda} \log\Bigg(1 - \frac{d}{2c+d}\Bigg) + \frac{1}{2} + \int_0^1 \frac{1}{2\Lambda} \log\Bigg(1 - \frac{d}{2c+d} \exp(-2\Lambda x)\Bigg) \textrm{d}x.
	\end{aligned}
\]

Let \(\operatorname{Li}_2\) denote the dilogarithm, \textit{i.e.}, the function defined as
\[
	\operatorname{Li}_2: x
	\; \longmapsto \;
	\int_0^x \frac{\log(1-u)}{u} \textrm{d}u.
\]
Thus,
\[
	\int_0^1 \log\Bigg(1 - \frac{d}{2c+d} \exp(-2\Lambda x)\Bigg) \textrm{d}x
	\; = \;
	\Bigg[-\frac{1}{2\Lambda} \operatorname{Li}_2\Bigg(\frac{d}{2c+d} \exp(-2\Lambda x)\Bigg)\Bigg]_0^1.
\]

\noindent\textbf{Assemblage.} Grouping all of our results, the second moment of the admixture fraction \(H_{\textcolor{vermillion}{S_1}}\) is
\[
	\begin{aligned}
		\mathbf{E}\big[H_{\textcolor{vermillion}{S_1}}^2\big]
		& \; = \;
		\frac{b}{2d} + \frac{ad - bc}{d} \frac{2}{2c+d} \Bigg(\frac{1}{2} - \frac{1}{2\Lambda} \log\Bigg(1-\frac{d}{2c+d}\Bigg) + \frac{1}{2\Lambda} \Bigg[-\frac{1}{2\Lambda} \operatorname{Li}_2\Bigg(\frac{d}{2c+d} \exp(-2\Lambda x)\Bigg)\Bigg]_0^1 \Bigg)
		\\
		& \; = \;
		\frac{b}{2d} + \frac{ad - bc}{d(2c+d)} \Bigg(1 + \frac{1}{\Lambda} \log\Bigg(\frac{2c+d}{2c}\Bigg) + \frac{1}{2\Lambda^2} \Bigg(\operatorname{Li}_2\Bigg(\frac{d}{2c+d} \exp(-2\Lambda)\Bigg) - \operatorname{Li}_2\Bigg(\frac{d}{2c+d}\Bigg)\Bigg)\Bigg).
	\end{aligned}
\]
To obtain the variance, we simply need to subtract the squared expectation. Note that
\[
	1 - h^2 + \frac{h^2}{N} - \frac{h^4}{N}
	\; = \;
	(1-h) (1+h) \Bigg(1+\frac{h^2}{N}\Bigg),
\]
so that
\[
	\frac{b}{2d}
	\; = \;
	\frac{2h \Big(1+\frac{h^2}{N}\Big) \textcolor{vermillion}{s_1}^2 \Big(\frac{1+h}{1-h}\Big)}{2h \Big(1+\frac{h^2}{N}\Big) (1-h) (1+h)}
	\; = \;
	\frac{\textcolor{vermillion}{s_1}^2}{(1-h)^2}
	\; = \;
	\mathbf{E}\big[H_{\textcolor{vermillion}{S_1}}\big]^2.
\]
Therefore,
\[
	\mathbf{V}\big[H_{\textcolor{vermillion}{S_1}}\big]
	\; = \;
	\frac{ad - bc}{d} \frac{1}{2c+d} \Bigg(1 + \frac{1}{\Lambda} \log\Bigg(\frac{2c+d}{2c}\Bigg) + \frac{1}{2\Lambda^2} \Bigg(\operatorname{Li}_2\Bigg(\frac{d}{2c+d} \exp(-2\Lambda)\Bigg) - \operatorname{Li}_2\Bigg(\frac{d}{2c+d}\Bigg)\Bigg)\Bigg).
\]
To simplify further,
\[
	\begin{aligned}
		2c
		\; = \;
		\frac{(1-h) C}{N},
		\quad \textrm{where }
		C
		& \; = \;
		2N \Bigg(1 - h^2 + \frac{h^2}{N} - \frac{h^4}{2N}\Bigg)
		\\
		& \; = \;
		2N \big(1-h^2\big) + h^2 \big(2-h^2\big),
	\end{aligned}
\]
and hence
\[
	\begin{aligned}
		2c+d
		\; = \;
		\frac{(1-h) B}{N},
		\quad \textrm{where }
		B
		& \; = \;
		C + h(1+h) \big(N + h^2\big)
		\\
		& \; = \;
		2N \big(1-h^2\big) + N h (1+h) + h^2 \big(2-h^2\big) + h^3 (1-h).
		\\
		& \; = \;
		N (1+h) (2-h) + h^2 (2+h).
	\end{aligned}
\]
In particular,
\[
	\frac{2c+d}{2c}
	\; = \;
	\frac{B}{C}.
\]
On the one hand,
\[
	\begin{aligned}
		ad-bc
		& \; = \;
		\Bigg(\Bigg(1 + \frac{h^2}{N}\Bigg) (1+h) \textcolor{vermillion}{s_1}^2 + \Bigg(1 - h^2 + \frac{h^2}{N}\Bigg) \textcolor{vermillion}{s_1}\Bigg) h (1-h) (1+h) \Bigg(1 + \frac{h^2}{N}\Bigg)
		\\ & \phantom{\; = \;}
		- 2h \Bigg(1 + \frac{h^2}{N}\Bigg) \Bigg(\frac{1+h}{1-h}\Bigg) \textcolor{vermillion}{s_1}^2
		(1-h) \Bigg(1 - h^2 + \frac{h^2}{N} - \frac{h^4}{2N} \Bigg)
		\\
		& \; = \;
		h (1+h) \Bigg(1 + \frac{h^2}{N}\Bigg) \textcolor{vermillion}{s_1} \Bigg( \Bigg(1 + \frac{h^2}{N}\Bigg) (1+h) \textcolor{vermillion}{s_1} (1-h) + \Bigg(1 - h^2 + \frac{h^2}{N} \Bigg) (1-h) - 2 \textcolor{vermillion}{s_1} \Bigg( 1 - h^2 + \frac{h^2}{N} + \frac{h^4}{2N} \Bigg) \Bigg)
		\\
		& \; = \;
		h (1+h) \Bigg(1 + \frac{h^2}{N}\Bigg) \textcolor{vermillion}{s_1} \Bigg( \Bigg(1 - h^2 + \frac{h^2}{N} \Bigg) (1-h) + \Bigg( 1 - h^2 + \frac{h^2}{N} + \frac{h^4}{N} \Bigg) \textcolor{vermillion}{s_1} - \Bigg( 2 - 2h^2 + \frac{2h^2}{N} + \frac{h^4}{N} \Bigg) \textcolor{vermillion}{s_1} \Bigg)
		\\
		& \; = \;
		h (1+h) \Bigg(1 + \frac{h^2}{N}\Bigg) \textcolor{vermillion}{s_1} \Bigg(1 - h^2 + \frac{h^2}{N}\Bigg) \big(1 - h - \textcolor{vermillion}{s_1}\big).
	\end{aligned}
\]
Recall that \(1 - h = s\) and \(1 - h - \textcolor{vermillion}{s_1} = \textcolor{blue}{s_2}\). Thus,
\[
	\begin{aligned}
		\frac{ad-bc}{d(2c+d)}
		& \; = \;
		\frac{h (1+h) \Big(1 + \frac{h^2}{N}\Big) \Big(1 - h^2 + \frac{h^2}{N}\Big) \textcolor{vermillion}{s_1} \big(1 - h - \textcolor{vermillion}{s_1}\big)}{h (1-h) (1+h) \Big(1 + \frac{h^2}{N}\Big) \Big(\frac{(1-h) B}{N}\Big)}
		\\
		& \; = \;
		\frac{\big(N - Nh^2 + h^2\big) \textcolor{vermillion}{s_1} \textcolor{blue}{s_2}}{s^2 B}
		\\
		& \; = \;
		\frac{\textcolor{vermillion}{s_1} \textcolor{blue}{s_2}}{s^2} \frac{A}{B},
		\quad \textrm{where }
		A
		\; = \;
		N \big(1 - h^2\big) + h^2.
	\end{aligned}
\]
On the other hand,
\[
	\begin{aligned}
		\frac{d}{2c+d}
		\; = \;
		1 - \frac{2c}{2c+d}
		\; = \;
		1 - \frac{C}{B}
		\; = \;
		\frac{D}{B},
		\quad \textrm{where }
		D
		& \; = \;
		B - C
		\\
		& \; = \;
		h (1+h) \big(N + h^2\big).
	\end{aligned}
\]
With all these definitions,
\[
	\mathbf{V}\big[H_{\textcolor{vermillion}{S_1}}\big]
	\; = \;
	\frac{\textcolor{vermillion}{s_1} \textcolor{blue}{s_2}}{s^2} \frac{A}{B} \Bigg(1 + \frac{1}{\Lambda} \log\Bigg(\frac{B}{C}\Bigg) + \frac{1}{2\Lambda^2} \Bigg(\operatorname{Li}_2\Bigg(\frac{D}{B} \exp(-2\Lambda)\Bigg) - \operatorname{Li}_2\Bigg(\frac{D}{B}\Bigg)\Bigg)\Bigg).
\]

\noindent\textbf{Limiting cases.} In the model of \cite{verduGeneralMechanisticModel2011}, both the population and the genome are assumed to be sufficiently large to ignore all linkage disequilibrium, whether due to limited recombination or genetic drift, can be neglected. Under ongoing gene flow, they find that
\[
	\mathbf{V}\big[\widetilde{H}_{\textcolor{vermillion}{S_1}}\big]
	\; = \;
	\frac{\textcolor{vermillion}{s_1}\textcolor{blue}{s_2}}{s(1+s)}
\]
where \(\widetilde{H}_{\textcolor{vermillion}{S_1}}\) denotes the admixture fraction under their model, as opposed to \(H_{\textcolor{vermillion}{S_1}}\) under ours.

To recover their regime, we compute the stationary variance \(\mathbf{V}\big[H_{\textcolor{vermillion}{S_1}}\big]\) for a large population (\(N \to \infty\)) and long chromosomes (\(\Lambda \to \infty\)). We consider only one order of limits, taking \(\Lambda \to \infty\) first and then \(N \to \infty\). Reversing the order yields the same limiting value, as can be verified directly from the expression above.

Since \(A\), \(B\), \(C\) and \(D\) do not depend on \(\Lambda\),
\[
	\begin{aligned}
		\lim_{N \to \infty} \lim_{\Lambda \to \infty}  \mathbf{V}\big[H_{\textcolor{vermillion}{S_1}}\big]
		& \; = \; \lim_{N \to \infty} \frac{\textcolor{vermillion}{s_1} \textcolor{blue}{s_2}}{s^2} \frac{A}{B}
		\\
		& \; = \;
		\lim_{N \to \infty} \frac{\textcolor{vermillion}{s_1} \textcolor{blue}{s_2}}{s^2} \frac{N (1-h^2) + h^2}{N (1+h) (2-h) + h^2 (2+h)}
		\\
		& \; = \;
		\frac{\textcolor{vermillion}{s_1} \textcolor{blue}{s_2}}{s^2} \frac{1-h^2}{(1+h)(2-h)}
		\\
		& \; = \;
		\frac{\textcolor{vermillion}{s_1} \textcolor{blue}{s_2}}{s^2} \frac{1-h}{2-h}
		\\
		& \; = \;
		\frac{\textcolor{vermillion}{s_1} \textcolor{blue}{s_2}}{s(1+s)}.
	\end{aligned}
\]
Thus, at least with respect to the stationary variance of admixture fractions, the model of \cite{verduGeneralMechanisticModel2011} is recovered as a limiting case of ours.

Another limiting case is the absence of recombination, that is, \(\Lambda \to 0\). Differentiating in \(\Lambda\) gives
\[
	\begin{aligned}
		\frac{\textrm{d}}{\textrm{d}\Lambda} \operatorname{Li}_2\Bigg(\frac{D}{B} \exp(-2\Lambda)\Bigg)
		& \; = \;
		2 \log\Bigg(1 - \frac{D}{B} \exp(-2\Lambda)\Bigg),
		\\
		\frac{\textrm{d}^2}{\textrm{d}\Lambda^2} \operatorname{Li}_2\Bigg(\frac{D}{B} \exp(-2\Lambda)\Bigg)
		& \; = \;
		\frac{4 \frac{D}{B} \exp(-2\Lambda)}{1 - \frac{D}{B} \exp(-2\Lambda)}.
	\end{aligned}
\]
Evaluating at \(\Lambda = 0\) yiels the following expansion:
\[
	\begin{aligned}
		\operatorname{Li}_2\Bigg(\frac{D}{B} \exp(-2\Lambda)\Bigg)
		& \; = \;
		\operatorname{Li}_2\Bigg(\frac{D}{B}\Bigg) + 
		2 \log\Bigg(1 - \frac{D}{B}\Bigg) \Lambda + \frac{4 \frac{D}{B}}{1 - \frac{D}{B}} \frac{\Lambda^2}{2} + \underset{\Lambda \to 0}{O}\big(\Lambda^3\big)
		\\
		& \; = \;
		\operatorname{Li}_2\Bigg(\frac{D}{B}\Bigg) + 2 \log\Bigg(\frac{C}{B}\Bigg) \Lambda + \frac{2D}{C} \Lambda^2 + \underset{\Lambda \to 0}{O}\big(\Lambda^3\big)
	\end{aligned}
\]
where we recall that \(C = B-D\). Therefore,
\[
	1 + \frac{1}{\Lambda} \log\Bigg(\frac{B}{C}\Bigg) + \frac{1}{2\Lambda^2} \Bigg(\operatorname{Li}_2\Bigg(\frac{D}{B} \exp(-2\Lambda)\Bigg) - \operatorname{Li}_2\Bigg(\frac{D}{B}\Bigg)\Bigg)
	\; = \;
	1 + \frac{D}{C} + \underset{\Lambda \to 0}{O}\big(\Lambda\big),
\]
which gives
\[
	\begin{aligned}
		\lim_{\Lambda \to \infty} \mathbf{V}\big[H_{\textcolor{vermillion}{S_1}}\big]
		& \; = \;
		\frac{\textcolor{vermillion}{s_1} \textcolor{blue}{s_2}}{s^2} \frac{A}{B} \Bigg(1 + \frac{D}{C}\Bigg)
		\\
		& \; = \;
		\frac{\textcolor{vermillion}{s_1} \textcolor{blue}{s_2}}{s^2} \frac{A}{C+D} \frac{C+D}{C}
		\\
		& \; = \;
		\frac{\textcolor{vermillion}{s_1} \textcolor{blue}{s_2}}{s^2} \frac{A}{C}
		\\
		& \; = \;
		\frac{\textcolor{vermillion}{s_1} \textcolor{blue}{s_2}}{2s^2} \frac{2N (1-h^2) + 2h^2}{2N (1-h^2) + 2h^2 + h^3}.
	\end{aligned}
\]
The first factor corresponds to the variance of the mean of two Bernoulli random variables, representing the ancestry of two non-recombining homologous chromosomes. The second factor is a correction term due to genetic drift, which slightly reduces this variance. As \(N \to \infty\), this correction term disappears as the two chromosomes become indepedent. Note that it also vanishes as \(h \to 0\), as genetic drift only occurs in the admixed population.

\end{document}